\documentclass[superscriptaddress,amsmath,amssymb,twocolumn,pra,floatfix,footinbib]{revtex4-1}
\usepackage[pdftex]{graphicx}
\usepackage{amsmath,amssymb}
\usepackage[usenames]{color}
\usepackage{mathbbol}
\usepackage{amssymb}
\usepackage{grffile}
\usepackage[breaklinks=true,colorlinks=true,linkcolor=blue,urlcolor=blue,citecolor=blue]{hyperref}
\usepackage{amsmath, amstext, amssymb, amsfonts, amsxtra}
\usepackage{textcomp}
\usepackage{xspace}
\usepackage{enumitem}
\usepackage{bbold}
\usepackage[normalem]{ulem}
\usepackage[section]{placeins}

\newcommand{\Rec}{\mathcal{R}}

\newcommand{\ket}[1]{|\!\! #1 \rangle}
\newcommand{\bra}[1]{\langle #1 \!\!|}

\newcommand{\rhop}{\hat{\rho}}

\newcommand{\figref}[1]{Fig.\ref{#1}}
\newcommand{\Ham}{\hat{\mathcal{H}}}

\newcommand{\sx}{\hat{\sigma}^x}
\newcommand{\sy}{\hat{\sigma}^y}
\newcommand{\sz}{\hat{\sigma}^z}

\begin{document}

\title{Transport and spectral properties of the XX$+$XXZ diode and stability to dephasing} 

\author{Kang Hao Lee}
\affiliation{Science and Math Cluster, Singapore University of Technology and Design, 8 Somapah Road, 487372 Singapore}

\author{Vinitha Balachandran}
\email{vinitha\_balachandran@sutd.edu.sg}
\affiliation{Science and Math Cluster, Singapore University of Technology and Design, 8 Somapah Road, 487372 Singapore}

\author{Chu Guo}
\affiliation{Henan Key Laboratory of Quantum Information and Cryptography, Zhengzhou,Henan 450000, China}
\affiliation{Key Laboratory of Low-Dimensional Quantum Structures and Quantum Control of Ministry of Education, Department of Physics and Synergetic Innovation Center for Quantum Effects and Applications, Hunan Normal University, Changsha 410081, China}

\author{Dario Poletti}
\email{dario\_poletti@sutd.edu.sg}
\affiliation{Science and Math Cluster, Singapore University of Technology and Design, 8 Somapah Road, 487372 Singapore}
\affiliation{Engineering Product Development Pillar, Singapore University of Technology and Design, 8 Somapah Road, 487372 Singapore}

\date{\today}

\begin{abstract}
We study the transport and spectral property of a segmented diode formed by an XX $+$ XXZ spin chain. This system has been shown to become an ideal rectifier for spin current for large enough anisotropy. Here we show numerical evidence that the system in reverse bias has signatures pointing towards the existence of three different transport regimes depending on the value of the anisotropy: ballistic, diffusive and insulating. In forward bias we observe two regimes, ballistic and diffusive. The system in forward and reverse bias shows significantly different spectral properties, with distribution of rapidities converging towards different functions. In the presence of dephasing the system becomes diffusive, rectification is significantly reduced, the relaxation gap increases and the spectral properties in forward and reverse bias tend to converge. For large dephasing the relaxation gap decreases again as a result of quantum Zeno physics.                                   
\end{abstract}

\maketitle

\section{\label{sec:level1}Introduction} 

The study of transport at the nanoscale has raised interest both for fundamental and applied reasons \cite{BertiniZnidaric2020, landi2021nonequilibrium,BenentiWhitney2017}. In particular, thanks to many-body interactions a dissipatively boundary driven system can turn from ballistic to diffusive and even insulating, as shown for the XXZ chain \cite{Prosen2011, BertiniZnidaric2020, landi2021nonequilibrium, BenentiRossini2009}. 
We note that spin chains with tunable interactions of lengths as large as $256$ spins have been realized using quantum simulators \cite{ebadi2021quantum} and that specifically the XXZ chain was implemented in \cite{scholl2021microwave}. Furthermore, different transport regimes of XXZ chain were also demonstrated using ultracold atoms of $^{7}$Li in \cite{Jepsennature2020}.  
%
%
Recently there has been particular interest in studying the phenomenon of rectification in strongly interacting spin chain, both for spin and heat currents \cite{PhysRevLett.126.077203,poulsen2021entanglement,Balachandran2018, Balachandran2019, LeePoletti2019, lee2021giant}. A setup of particular interest is the XX + XXZ chain, composed, for half, of an XX, and the other half, of an XXZ (semi-)chain \cite{BiellaFazio2016,BiellaMazza2019}. 
With this setup, it was shown that one can achieve perfect rectification in the thermodynamic limit \cite{Balachandran2018}. This has been explained by the emergence of an insulating phase for large enough anisotropy/interaction in reverse bias, while the system still allows transport in forward bias. 
Hence, the emergence of significantly different transport properties in this system can be linked to an out-of-equilibrium phase transition. Phase transitions in dissipative systems have been studied with greater effort in the past ten years \cite{Kessler2012, Minganti2018, morrison2008a} even with experimental setups \cite{FitzpatrickHouck2017}. In models described by master equations in Gorini-Kossakowski-Sudarshan-Lindblad  (GKSL) form \cite{Lindblad1976,Gorini1976}, one can recognize the emergence of a dissipative phase transition by studying the rapidities, i.e. the rate of relaxation of the natural modes of the system. It has been shown that, at the transition, the relaxation gap closes, i.e. more than one mode has a vanishing relaxation rate \cite{Kessler2012, Minganti2018}. This however does not apply to systems for which the dissipation only acts at the boundaries, because in these systems the effects of the dissipation become a small perturbation in the thermodynamic limit \cite{Znidaric2015}. In these cases, in the thermodynamic limit the relaxation gap always closes, but the scaling of the smallest rapidity with the system size varies, for example from algebraic to exponential, or between two different algebraic decays at the phase transition \cite{Znidaric2015}. 

\begin{figure}[!htb]
    \includegraphics[trim={0 0 0.1cm 0},clip,width=\columnwidth]{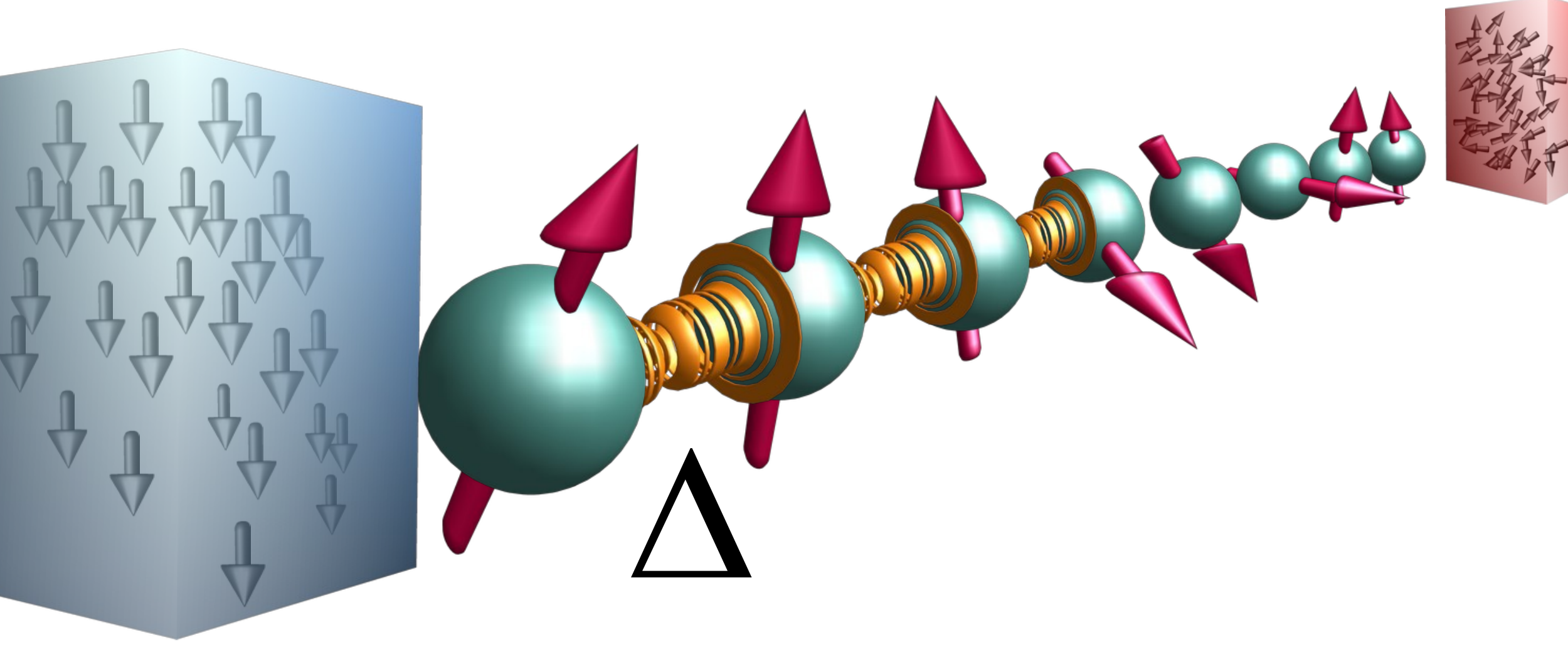}
    \caption{The boundary driven segmented XXZ spin rectifier with anisotropy $\Delta$ only on the left half of the system. The diagram above represents the reverse bias scenario where the left bath tends to set the left most spin pointing down and the right bath tends to set the right most spin to a state which is an equal mixture of up and down spins.}
    \label{fig:xxzpic}
\end{figure} 

In this work we achieve a deeper understanding of the transport and spectral properties of XX $+$ XXZ spin diode depicted in Fig.\ref{fig:xxzpic} and studied in \cite{Balachandran2018}. Our studies are numerical and performed with exact diagonalization and tensor-networks based algorithms \cite{White1992, Schollwock2011}. With the parameters considered, both the approaches are limited in system size and this poses limitations to characterize the system in the thermodynamic limit.
Our transport studies suggest that in forward bias the system is ballistic for lower interactions, and diffusive for larger ones. Furthermore, in reverse bias the system is ballistic, diffusive or insulating as one increases the interaction. Our spectral analysis shows that the rapidities tend to different distributions in forward and reverse bias. Moreover, a qualitative change in the scaling of the relaxation gap of the rapidities is seen across the different transport regimes.
Last, we will also add to the boundary dissipation that drives the system, dephasing at each site, something which affects significantly both the transport properties of the system, and the distribution of the rapidities. Indeed, with dephasing we observe the increase of relaxation gap as well as significant reduction of rectification along with the convergence of spectral properties of the forward and reverse bias. With very large dephasing, the relaxation gap decreases again due to the quantum Zeno effect.

This work is organized in the following manner. In Sec.\ref{sec:model} we describe the model. In Sec.\ref{sec:currents} we consider the currents, in Sec.\ref{sec:magnetization} we describe the magnetization, and in Sec.\ref{sec:spectr} the spectral properties. Finally in Sec.\ref{sec:dephasing} we discuss the effect of dephasing and we give our conclusions in Sec.\ref{sec:conclusions}.  

\section{Model}\label{sec:model}

In the following we consider an even number $L$ of spins in a bipartite spin-1/2 chain described by the Hamiltonian in \cite{Balachandran2018} (pictorial representation in \figref{fig:xxzpic})
\begin{align}
\Ham &= \sum_{n=1}^{L/2-1} [J (\sx_n\sx_{n+1} + \sy_n\sy_{n+1} + \Delta\; \sz_n\sz_{n+1})] \nonumber\\
&+ \sum^{L-1}_{n=L/2+1} [J (\sx_n\sx_{n+1} + \sy_n\sy_{n+1})] \nonumber \\
&+J_m (\sx_{L/2}\sx_{L/2+1} + \sy_{L/2}\sy_{L/2+1}).\label{baseHam}
\end{align}
$\sx_n$, $\sy_n$ and $\sz_n$ are the Pauli matrices for the $n$-th spin, $J$ is the tunnelling amplitude, $\Delta$ is the anisotropy in the left half of the system, and $J_m$ is the tunnelling amplitude between the two half chains. The left half chain thus follows the XXZ model and the right half of the chain follows the XX model. 

The chain is coupled to two different spin baths at the boundaries and also subjected to bulk dephasing. We model the effect of the baths via the following master equation in GKSL  form  \cite{Lindblad1976,Gorini1976}
\begin{align}
\frac{d\hat{\rho}}{d t} = \mathcal{L}(\rho) &\nonumber \\ 
=-\frac{\rm{i}}{\hbar}[\hat{\mathcal{H}},\hat{\rho}] & +\sum^{4}_{n=1}B_n\hat{\rho} B_n^\dagger-\frac{1}{2}\sum^{4}_{n=1}\{B^\dagger_n B_n,\hat{\rho}\},\nonumber \\ 
& + \sum^{L}_{n=1}D_n\hat{\rho} D_n^\dagger-\frac{1}{2}\sum^{L}_{n=1}\{D^\dagger_nD_n,\hat{\rho}\}    
\label{eq:masterequation}
\end{align}
where the spin polarization processes at the boundaries are given by,
\begin{equation}
\begin{aligned}
    B_{1} &= \sqrt{\Gamma\mu_l}\hat{\sigma}_1^+, & B_{2} &=\sqrt{\Gamma(1-\mu_l)}\hat{\sigma}_1^-,\\
    B_{3} &= \sqrt{\Gamma\mu_r}\hat{\sigma}_L^+, & B_{4} &=\sqrt{\Gamma(1-\mu_r)}\hat{\sigma}_L^-,
\end{aligned}
\label{eq: set of dissipators}
\end{equation}
and the bulk dephasing of all spins,
\begin{equation}
\begin{aligned}
    D_n &= \sqrt{\gamma}\sz_n, & n\in\{1,...,L\};
\end{aligned}
\label{eq:dephasing}
\end{equation}
$\gamma$ controls the rate of dephasing, $\Gamma$ is the magnitude of the boundary system-bath coupling and $\mu_l$ ($\mu_r$) sets the spin magnetization imposed by the left (right) bath. Here we point out that, in the absence of dephasing, and for $\mu_l=0$ and $\mu_r=1$, an XX chain is always ballistic, while an XXZ chain would be ballistic for $\Delta<\Delta_{XXZ}$ and insulating for $\Delta>\Delta_{XXZ}$ where $\Delta_{XXZ}=1$.       

Here, we set $\mu_l$ and $\mu_r$ to be equal to 0 or 0.5. A spin current is induced when $\mu_l$ is different from  $\mu_r$. When $\mu_l=0.5>\mu_r=0$, the left bath tends to set the spins to be in an equal mixture of up and down spins $({\ket{\downarrow}_n\bra{\downarrow}}+{\ket{\uparrow}_n\bra{\uparrow}})/2$ while the right bath tends to set the spins pointing down ${\ket{\downarrow}_n\bra{\downarrow}}$. Spin current thus flows from the left to right and we refer to this as the forward bias. 
The reverse bias corresponds to the spin current flowing from right to left ($\mu_l=0<\mu_r=0.5$). 

The spin current $\mathcal{J}$ is calculated from the continuity equation for local observable $ \hat{\sigma}_n^z$ and is given by
\begin{equation}
\mathcal{J} = \textrm{Tr}(\hat{j}_n \hat{\rho}_{ss}), \label{eq:steadycurrent}
\end{equation}
where $\hat{\rho}_{ss}$ is the steady state such that $d\hat{\rho}_{ss}/dt=0$ and
\begin{equation}
\begin{aligned}
\hat{j}_{n}  = \frac{2J}{\hbar}({\sx}_n {\sy}_{n+1} - {\sy}_n {\sx}_{n+1}).\\
\end{aligned}
\label{current}
\end{equation}
We note that the spin current $\mathcal{J}$ is site independent in the steady state, and for this reason $\mathcal{J}$ in Eq.(\ref{eq:steadycurrent}) has no index $n$. For the rest of the paper, we work in units such that $J=\hbar = 1$. 

The spin magnetization observable on site $n$ is given by
\begin{equation}
    \langle \hat{\sigma}_n^z \rangle =  \textrm{Tr}(\hat{\sigma}_n^z \;\hat{\rho_{ss}}),
\end{equation}
and their profiles are discussed in Sec.\ref{sec:magnetization}.
 
To evaluate the performance of the system as a diode, we consider the measure $\mathcal{R}$ \cite{Balachandran2018}
\begin{equation}
\mathcal{R}=-\frac{\mathcal{J}_f}{\mathcal{J}_r},
\label{eq: contrast}
\end{equation}
where $\mathcal{J}_f$ and $\mathcal{J}_r$ are the currents in forward and reverse bias respectively.
When there is no rectification, $\mathcal{J}_f=-\mathcal{J}_r$ and the measure $\mathcal{R}=1$. For a perfect diode, $\mathcal{R}= \infty$ or $0$ because either the forward or reverse current goes to zero. 
In \cite{Balachandran2018} it was shown that the reverse bias system becomes an insulator for $\Delta>\Delta_R$ where $\Delta_R=1+\sqrt{2}$. 

Here we would like to spend a few lines on how we compute the steady state and the spectrum of the GKSL master equation. To obtain the entirety of the spectrum we write the right hand side of Eq.(\ref{eq:masterequation})  as $\mathcal{L}(\rhop)$ where $\mathcal{L}(\cdot)$ is a linear non-Hermitian superoperator which for our case, once written as a matrix, can be diagonalized. The eigenvalues are referred to as {\it rapidities}. These are, typically, complex numbers, whose real part describes the decay of the eigenmode, while the imaginary part its oscillations. For the master equation we consider, there exist a single steady state, which corresponds to a rapidity equal to zero. It becomes apparent that this procedure to compute all the rapidities can only be performed for small systems because the size of the Hilbert space grows approximately as $2^{2L}$ \cite{footnote1}. Another very important issue regards the accuracy of these calculations when $\Delta$ and $L$ are larger. In fact, in this case there are a number of rapidities which approach zero, thus making the diagonalization procedure less accurate. If we do not require to obtain the full spectrum, we can use two other approaches. The first one is to use iterative diagonalization routines that only return the eigenvalue with smallest modulus, which has modulus equal to zero and its eigenmode corresponds to the steady state. The other approach is to evolve the system in time, which we actually implement with a matrix product states algorithm \cite{Schollwock2011}. However, even in these cases the problem of small rapidities still applies. For instance, one would need to know when to stop the time evolution. For this we measure the current at each bond and compute the variance. When the variance is small enough, we stop our calculations and study the steady state. However, when the steady state current is small, say of the order of $10^{-5}$ it is very difficult to have a small enough variance between the different bonds, especially with such small rapidities. In short this is to state that for the system and bath parameters considered, it is still challenging to study large systems. For this reason, in this work we do not go beyond systems of size $L=20$. 

\section{Spin currents}\label{sec:currents}  
To characterize the transport properties, we begin by studying the finite size scaling of the spin current. In some scenarios, significantly large system sizes are required for the correct prediction of thermodynamic transport properties. As explained above, our studies are restricted to fairly small sizes. We will thus give our considerations from the sizes we observe and from our understanding of the systems, laying the foundation for future research on these systems. 
\begin{figure}[!htb]
    \includegraphics[trim={0 0 0.1cm 0},clip,width=\columnwidth]{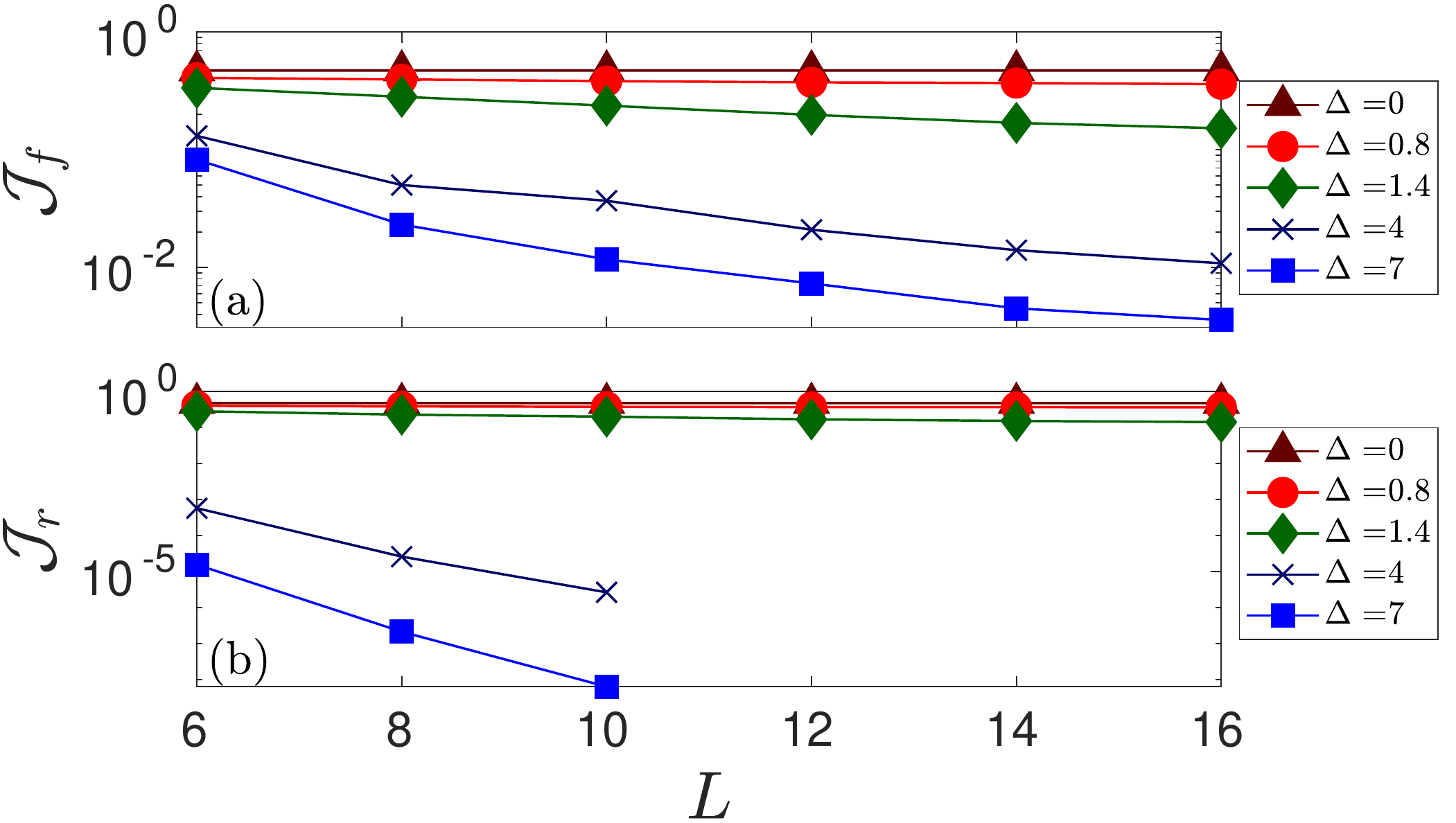}
    \caption{Forward current $\mathcal{J}_f$ panel (a) and reverse current $\mathcal{J}_r$ panel (b) versus the system size $L$ for selected anisotropy $\Delta$ values. Other parameters: $J_m=1$, $\Gamma=1$, $\gamma=0$.}
    \label{fig:jlvsdelta}
\end{figure} 

Fig.\ref{fig:jlvsdelta} shows the steady state spin current versus system sizes for  different anisotropies $\Delta$. In general, the current decays algebraically with the system size, i.e. $\mathcal{J}\propto1/L^\alpha$. When $\alpha=0$, the current is size independent and the transport is ballistic, while for $\alpha=1$ the system is diffusive. However, different values of $\alpha$ are also possible, and the system is categorized as superdiffusive for $0<\alpha<1$ and subdiffusive for $\alpha>1$. It is however important to note that the current may decay faster than algebraically with the system size, for example exponentially, in insulators. In Fig.\ref{fig:jlvsdelta}(a) we plot the currents in forward bias and we see two regimes of transport. For $\Delta \lesssim \Delta_{XXZ}$ the system is ballistic, similar to \cite{Prosen2011,ProsenIlievski2013}, while the system for $\Delta > \Delta_{XXZ}$ appears not to decay exponentially with the system size. 

The reverse bias currents are plotted in Fig.\ref{fig:jlvsdelta}(b), and they display a richer transport behavior with at least three clear regimes. The transport is ballistic for $\Delta < \Delta_{XXZ}$, whereas for $\Delta_{XXZ} < \Delta < \Delta_R $, the current appear to decay algebraically with the system size. Note that the magnitude of the current in reverse bias is significantly smaller compared to the forward bias case. Last, for $\Delta>\Delta_R$ the indication from our numerical calculations, and the insights developed in \cite{Balachandran2018}, indicate that the current decays exponentially with the system size. To help the reader differentiate the three regimes, in Fig.\ref{fig:jlvsdelta} we have used different colors for each one, specifically red for $\Delta \lesssim \Delta_{XXZ}$, green for $\Delta_{XXZ} < \Delta < \Delta_R $, and blue for $\Delta > \Delta_R $. Furthermore, within each region, darker shades correspond to smaller values. 
 
To gain deeper insight, we complement the analysis of the currents with that of the magnetization in the next section. 

\section{Spin Magnetization}\label{sec:magnetization} 
\begin{figure}[!htb]
    \includegraphics[trim={0 0 0.1cm 0},clip,width=\columnwidth]{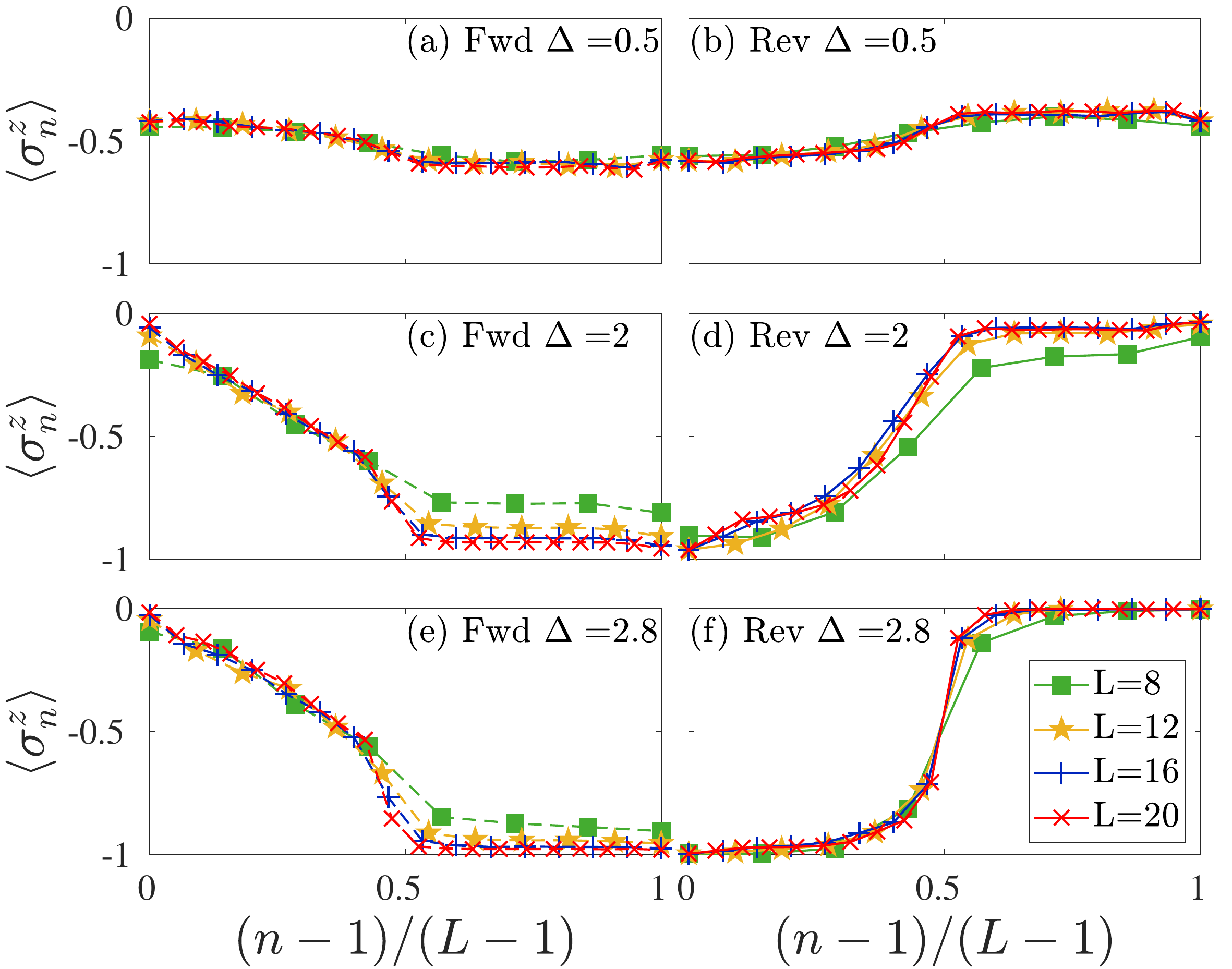}
    \caption{Spin magnetization profiles $\langle \hat{\sigma}_n^z \rangle$ against the spin positions (scaled) $(n-1)/(L-1)$ for $\Delta=0.5$ (a,b),  $\Delta=2$ (c,d), $\Delta=2.8$ (e,f)  in the forward (a,c,e) and reverse (b,d,f) directions. Other parameters: $\Gamma=1$, $J_m=1$, $\gamma=0$. }
    \label{fig:szprofile}
\end{figure} 

To establish the existence of distinct regimes seen in the transport in the previous section, we further investigate the spin magnetization in the two biases. First in Fig.\ref{fig:szprofile}, we analyze the system by studying its magnetization profile. We use a rescaled position $(n-1)/(L-1)$ so that for each system size the plot is between $0$ and $1$. In the left panels, i.e. Fig.\ref{fig:szprofile}(a,c,e), we show the profile for forward bias, while in the right panels, i.e. Fig.\ref{fig:szprofile}(b,d,f), we depict results from the reverse bias. In panels Fig.\ref{fig:szprofile}(a,b) we consider a small anisotropy $\Delta=0.5$, Fig.\ref{fig:szprofile}(c,d) an intermediate value $\Delta=2$ and in Fig.\ref{fig:szprofile}(e,f) a larger $\Delta=2.8$. In all scenarios we observe that the right portion of the chain, for $(n-1)/(L-1)>0.5$ the profile is approximately constant, and this is due to the fact that there the system is an $XX$ chain which is non-interacting and ballistic. The magnetization profiles are significantly different in the left portion of the chain where the anisotropy is non-zero. For both biases when $\Delta < \Delta_{XXZ}$, we observe that the magnetization at the edges of the chain is significantly far from the one the baths are trying to impose (either $-1$ or $0$), and most of the magnetization change occurs at the interface. In the left half of the chain we observe a gentle slope in the magnetization profile indicating possible diffusive or superdiffusive transport. 

The magnetization profile differs significantly from the forward and reverse bias once we consider larger values of $\Delta$.
In forward bias we observe that, for $\Delta > \Delta_{XXZ}$, Fig.\ref{fig:szprofile}(c,e), there is a distinct linear profile for the magnetization which would be expected from the diffusive nature of the XXZ chain \cite{Prosen2011b,ProsenZnidaric2012, landi2021nonequilibrium}, although in these studies the bias from the baths was different. At the interface we also observe a significant jump in the magnetization, which is a typical signature of interface resistance. Similar results are seen for reverse bias with anisotropy $\Delta_{XXZ}<\Delta<\Delta_R$ as in Fig.\ref{fig:szprofile}(d), for $\Delta=2$, although the magnetization slope is not a clear straight line as for the forward bias, which could be a hint of subdiffusive behavior. For large values of the anisotropy, $\Delta>\Delta_R$, shown in Fig.\ref{fig:szprofile}(f) a steep change of the reverse bias magnetization is seen which increases with the system size. Furthermore, the magnetization at the edge is extremely close to the values required by the baths. This is aligned with what we would expect for an insulator. 

\begin{figure}[!htb]
    \includegraphics[trim={0 0 0.1cm 0},clip,width=\columnwidth]{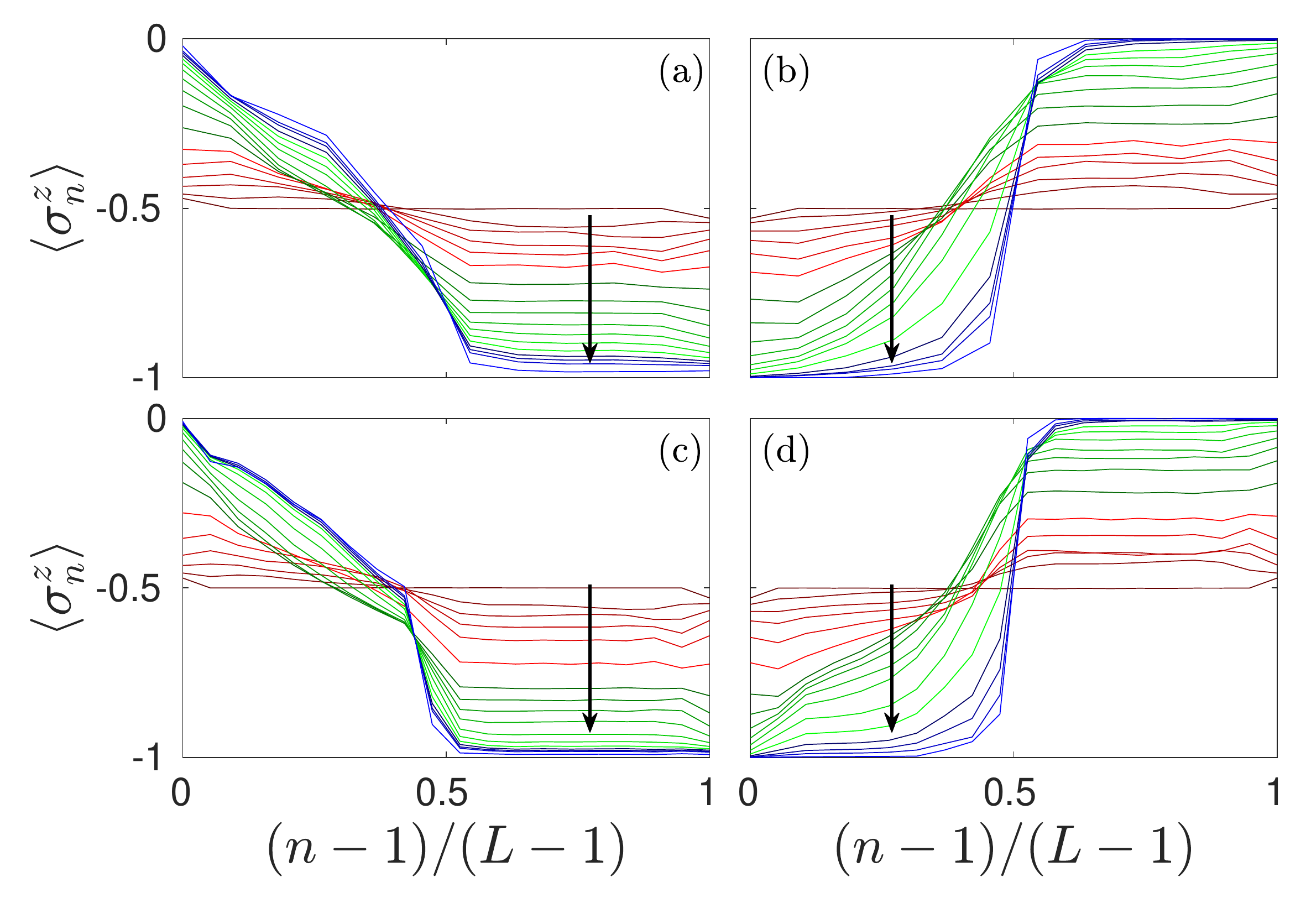}
    \caption{Panels (a-d): Spin magnetization profiles $\langle \hat{\sigma}_n^z \rangle$ against the spin positions (scaled) $(n-1)/(L-1)$ for $L=12$ (a,b) and $L=20$ (c,d) for different anisotropies $\Delta$ in the forward (a,c) and reverse (b,d) directions. Other parameters: $\Gamma=1$, $J_m=1$, $\gamma=0$. The arrow indicates increasing magnitudes of $\Delta$, specifically  $\Delta=$ [0 to 3 in intervals of 0.2, and $4$].}
    \label{fig:szprofile2}
\end{figure}

We further analyze the magnetization profiles in Fig.\ref{fig:szprofile2}, for two different sizes $L=12$ and $L=20$ for different values of the anisotropy $\Delta$. Specifically we use the values $\Delta=$ [0 to 3 in intervals of 0.2, and $4$] increasing in the direction of the arrows in Fig.\ref{fig:szprofile2}. 
To help differentiate the different regimes, we use the same coloring schemes as in Fig.\ref{fig:jlvsdelta}. The left panels of Fig.\ref{fig:szprofile2} correspond to forward bias, while the right panels to the reverse bias. Focusing on the forward bias, in Fig.\ref{fig:szprofile2}(a,c), consistently with Fig.\ref{fig:jlvsdelta}(a) and Fig.\ref{fig:szprofile}(a,c,e), there appears to be two main regimes: one for $\Delta < \Delta_{XXZ}$ for which the magnetization in the left chain is approximately constant, and for $\Delta > \Delta_{XXZ}$ for which the magnetization changes linearly until the interface. Furthermore, comparing the results for $L=12$, Fig.\ref{fig:szprofile2}(a), and for $L=20$, Fig.\ref{fig:szprofile2}(c), the two different regimes become more distinct, as the lines corresponding to different anisotropies becomes more clearly separated. In reverse bias, Fig.\ref{fig:szprofile2}(b,d), we also observe the three main regimes which also become more distinct as one compares the magnetization profile for $L=12$, Fig.\ref{fig:szprofile2}(b), and for $L=20$, Fig.\ref{fig:szprofile2}(d).      

 \begin{figure}[!htb]
    \includegraphics[trim={0 0 0.1cm 0},clip,width=\columnwidth]{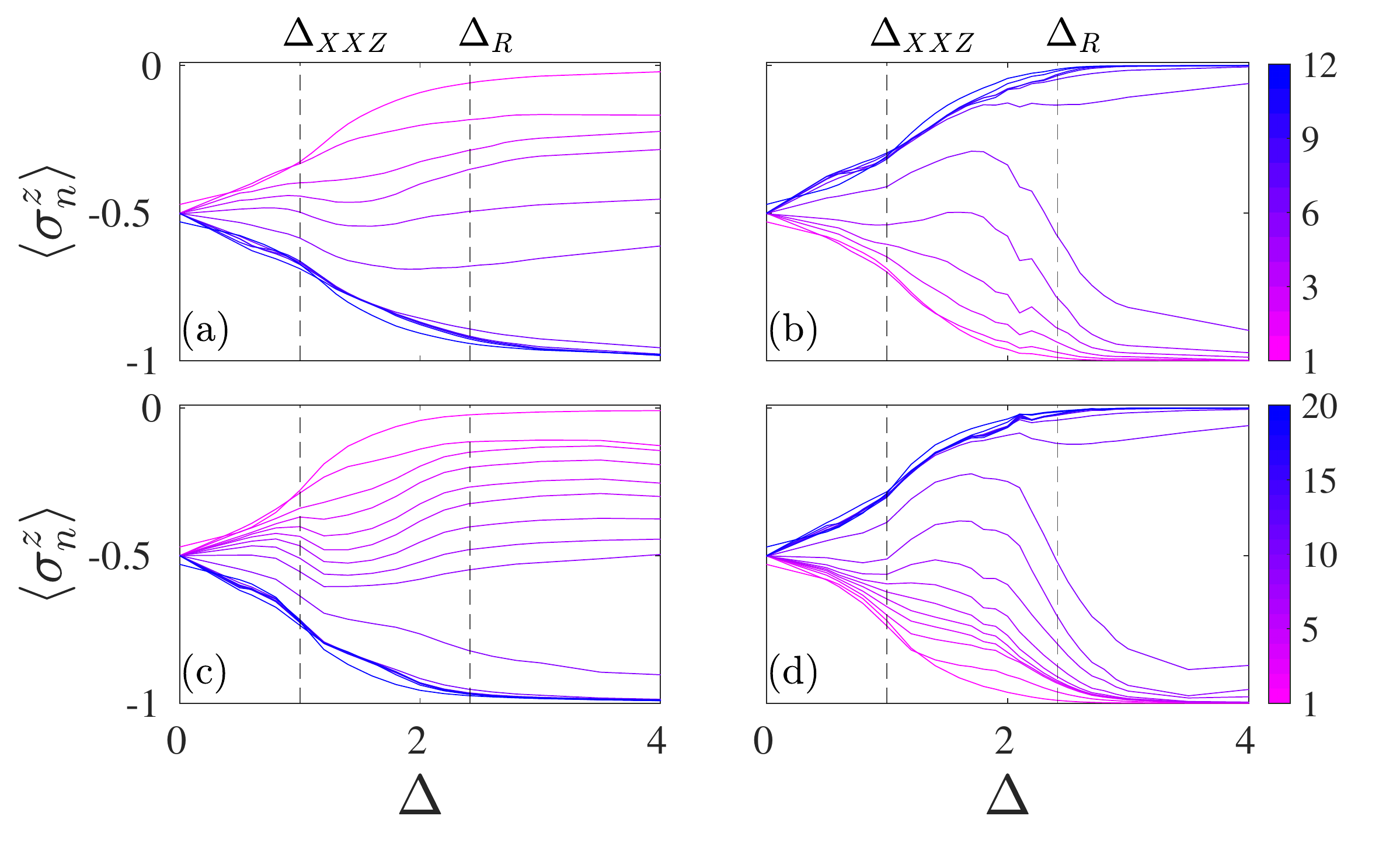}
    \caption{Spin magnetization profiles $\langle \hat{\sigma}_n^z \rangle$ versus $\Delta$ for different sites $n$. The case for $L=12$ is depicted in (a,b), and $L=20$ in (c,d), while the forward bias is represented in (a,c) and the reverse bias in (b,d). Other parameters: $\Gamma=1$, $J_m=1$, $\gamma=0$.}
    \label{fig:loc_magn}
\end{figure}

We then focus on a single site $n$, and plot in Fig.\ref{fig:loc_magn} the magnetization $\langle \sz_n \rangle$ versus the anisotropy $\Delta$. In this figure, each line corresponds to a different site (the color becomes darker as the site number increases, i.e. the lightest line corresponds to $n=1$, and the darkest one to $n=L$ as shown in the color bar). The forward bias case is considered in Fig.\ref{fig:loc_magn}(a,c) and reverse bias in Fig.\ref{fig:loc_magn}(b,d). In the reverse bias case we observe three different regions appearing at the values of $\Delta<\Delta_{XXZ}$, $\Delta_{XXZ}<\Delta<\Delta_R$ and $\Delta>\Delta_R$. The different behaviors in the different regions is much more marked for the larger system size $L=20$. In contrast, for the forward bias no significant change seems to occur at $\Delta=\Delta_R$, while a qualitative change in the shape of the magnetization versus interaction occurs at $\Delta = \Delta_{XXZ}$. This reinforces the observation that there are two main transport regimes in the forward bias scenario. 

Overall, our analysis of the current and of the magnetization profiles is consistent in pointing that there is a qualitative change in the properties of the system, in forward bias at $\Delta=\Delta_{XXZ}$, while in reverse bias there are two changes, one at $\Delta = \Delta_{XXZ}$ and one at $\Delta = \Delta_R$.

\section{Spectrum of rapidities}\label{sec:spectr}   

\begin{figure}[!htb]
            \includegraphics[width=\columnwidth]{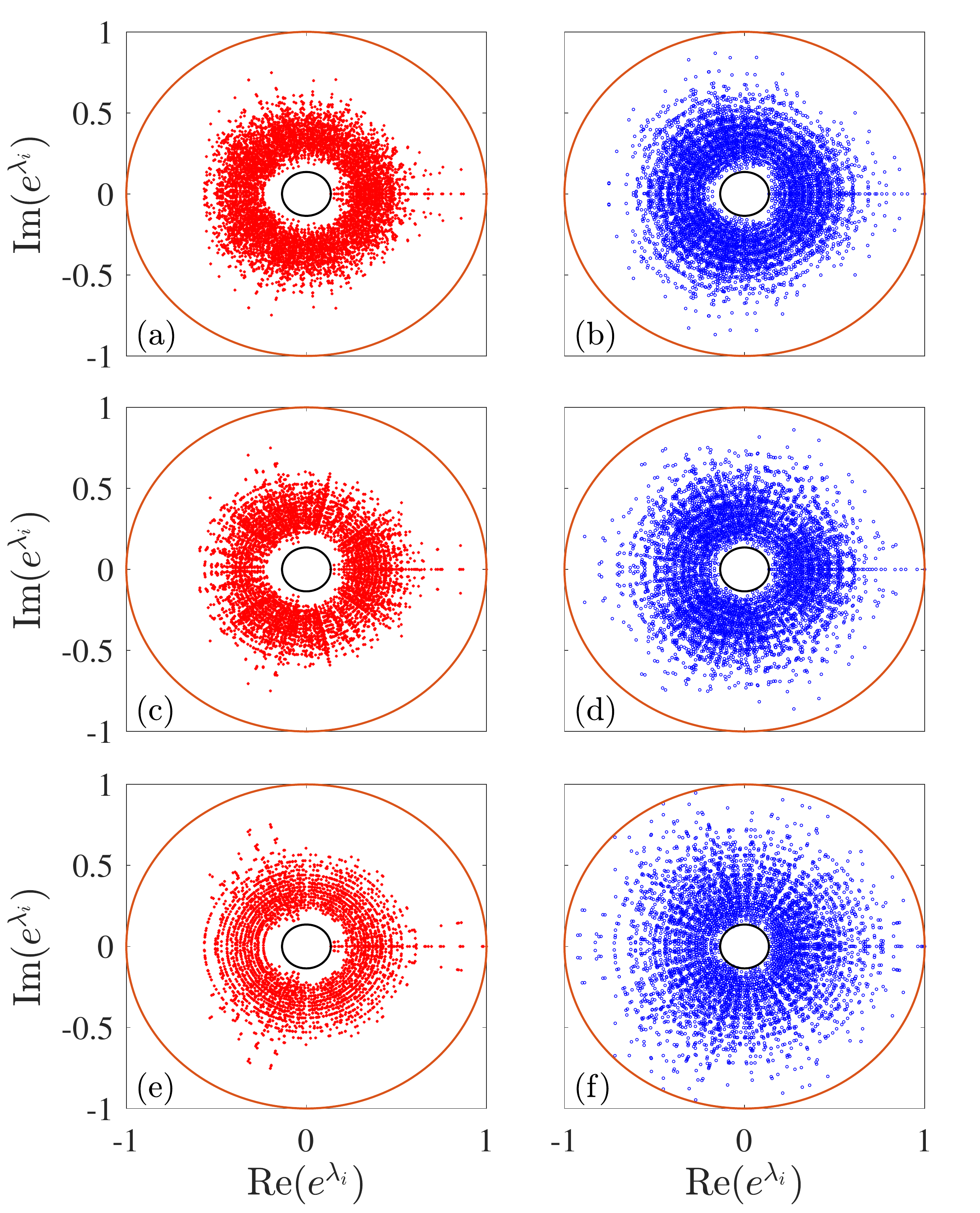}
    \caption{Exponential of the rapidities $\lambda_i$ in the unit circle (red continuous line) for anisotropies $\Delta=0.5$ panels (a,b), $\Delta=2$ (c,d), $\Delta=5$ (e,f). Left panels (a,c,e) are for forward bias and right panels (b,d,f) are for reverse bias. The black continuous inner circle shows the smallest possible radius of the exponential of rapidities \cite{Note1}. Other parameters: $L=8$, $\Gamma=1$, $J_m=1$, $\gamma=0$.}
    \label{fig:spectrum}
\end{figure} 

To investigate further the different properties of forward and reverse bias systems, we now turn our attention to the spectrum of the superoperator $\mathcal{L}$ from Eq.(\ref{eq:masterequation}). In fact, the spectrum of the superoperator plays an important role in determining the relaxation dynamics as well as the steady state properties of the open quantum system. One of the relevant quantities in this context is the spectral gap defined as $|Re(\lambda_1)|$ where $\lambda_1$ is the first non-zero eigenvalue of $\mathcal{L}$. The inverse of the spectral gap, also called the asymptotic decay rate, determines the slowest relaxation time scale in the dynamics of the system. In a system with dissipation acting on an extensive number of sites, the dissipative phase transitions is characterized by the closing of the gap, which results in the divergence of relaxation times and emergence of multiple steady states in the thermodynamic limit \cite{Kessler2012, Casteels2017,Horstmann2013,Minganti2018}. However, when the dissipation is at the boundaries only, as the case in our study, the gap always closes in the thermodynamic limit. Here, the phase transition is signalled by the change in the scaling of the gap \cite{Znidaric2015}.

In Fig.\ref{fig:spectrum} we analyze the spectrum by plotting the exponential of the rapidities $e^{\lambda_i}$, where the $\lambda_i$ are the rapidities in the complex plane. $e^{\lambda_i}$ is bound to be within the unit circle, with only one of them corresponding to the steady state with $\lambda=0$, on the unit circle \footnote{The maximum absolute value of the real part of the rapidity depends upon the number of the sites in the chain to which the baths are coupled. Hence in our setup the exponential of the rapidities fall within two concentric circles of radii $e^{-2}$ and $1$}. 
On the left panels, Fig.\ref{fig:spectrum}(a,c,e), we show the exponential of the rapidities for the forward bias, while the right panels, Fig.\ref{fig:spectrum}(b,d,f), depict the case of the reverse bias. We observe that for small $\Delta$ the rapidities are distributed similarly in the forward and reverse bias, but as $\Delta$ increases, in the reverse bias we see a number of rapidities approaching the unit circle closing the spectral gap, indicating the emergence of slowly decaying modes.     

\begin{figure}[!htb]
          \includegraphics[width=\columnwidth]{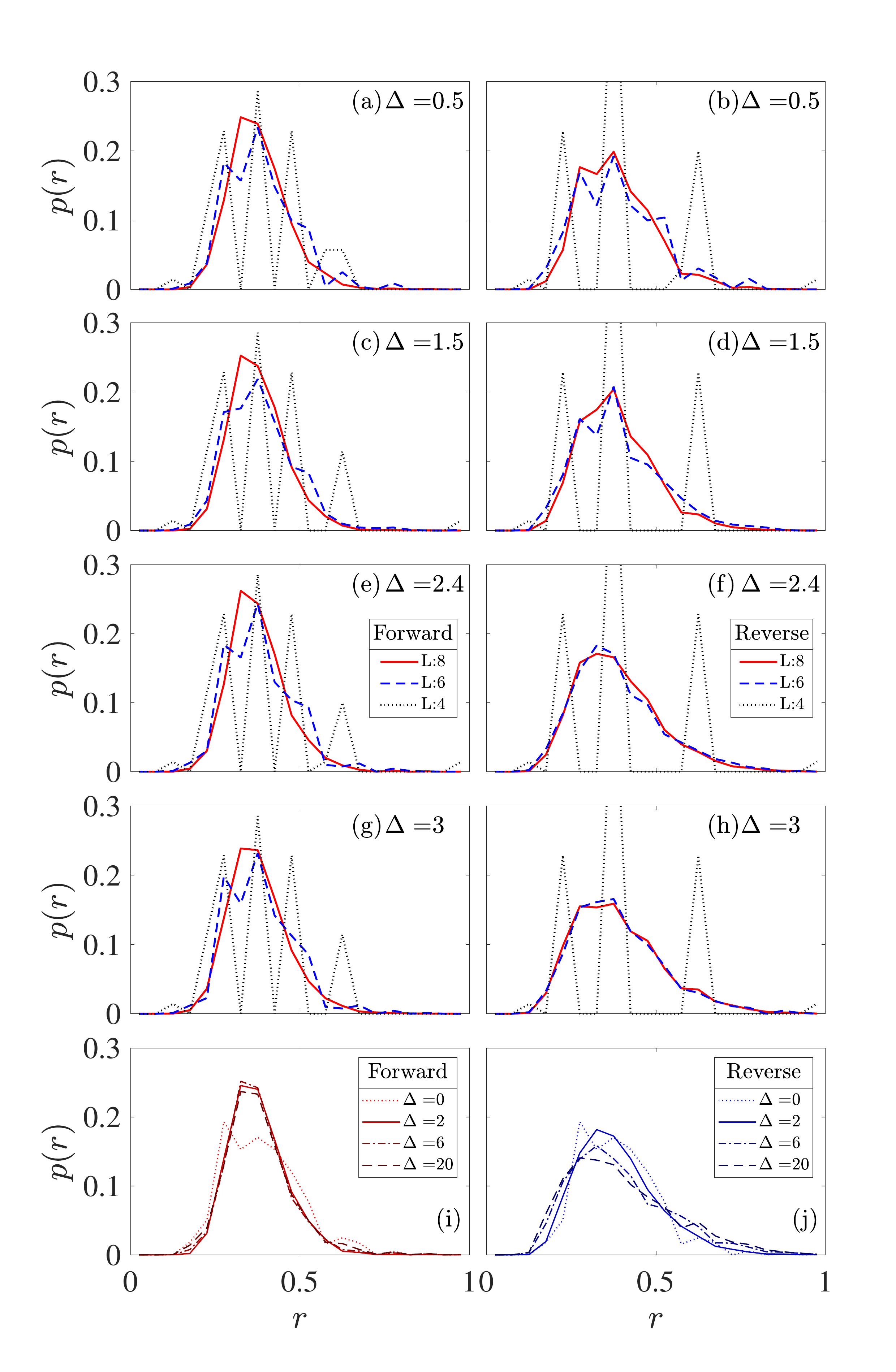}            
            \caption{Distribution of rapidities $p(r)$ against the radial distance $r$ in complex plane for  different anisotropies $\Delta=0.5$ panels (a,b), $\Delta=2.0$ (c,d) and $\Delta=2.8$ (e,f). In panels (g,h) distribution $p(r)$ is plotted against anisotropy $\Delta$ for system size $L=8$. Other parameters: $J_m=0.1$, $\Gamma=1$ and $\gamma=0$. Left panels are for forward bias and right panels are for reverse bias.}
            \label{fig:histrapidities}
\end{figure} 

To gain further insights we study the radial distribution of the rapidities by plotting the number of rapidities within concentric circles of radius $dr=0.05$ denoted by  $ p(r)$. These are depicted in \figref{fig:histrapidities}. In the left panels we consider the forward bias case, while the right panel shows the reverse bias. In panels (a-h) we consider the system size $L=4$ (black dotted), $L=6$ (blue dashed) and $L=8$ (red solid), and each panel for a different $\Delta$. For $L=4$ there are strong finite size effects, but already for $L=6$ and $L=8$ we see convergence in the distributions. Interestingly, we find that the distribution differs significantly in forward and in reverse bias. To see this more clearly, in panels Fig.\ref{fig:histrapidities}(i,j) we plot the distribution for $L=8$ for different anisotropies. For large values of anisotropy, forward and reverse bias distributions seems to tend towards a different, yet well defined, function. And the probability of rapidities close to $r=1$ is significantly larger for the reverse bias compared to the forward one, suggesting the presence of a significantly larger number of slow decaying modes, which is expected due to the much slower relaxation dynamics of the reverse bias system. The exact expression for these distributions can be a topic for future research. 

\begin{figure}[!htb]
            \includegraphics[width=\columnwidth]{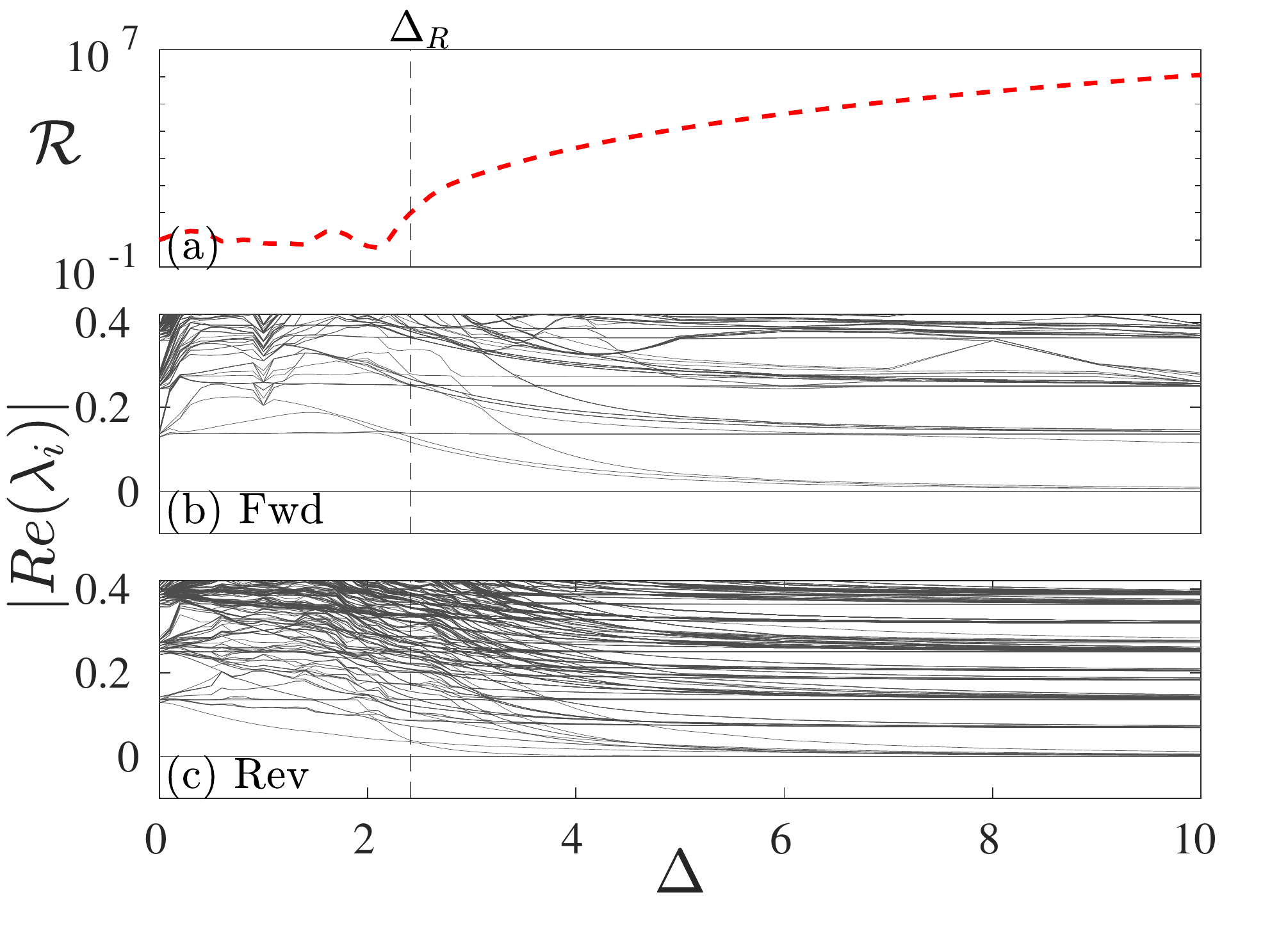}
    \caption{Panel (a): Rectification measure $\mathcal{R}$ as a function of anisotropy $\Delta$. Panels (b) and (c) shows $|Re(\lambda_i)|$ against $\Delta$ for forward and reverse direction respectively. Other parameters: $L=8$, $J_m=0.1$, $\Gamma=1$, $\gamma=0$. Vertical dashed line indicates $\Delta=\Delta_R$.} 
    \label{fig:gapvsjlzzR}
\end{figure} 

Since the occurrence of an out-of-equilibrium phase transition is related to the difference in the scaling of rapidities which converge to zero in the thermodynamic limit \cite{Znidaric2015}, we focus on the real part of the rapidities and plot it versus the anisotropy $\Delta$, both for forward, Fig.\ref{fig:gapvsjlzzR}(b), and reverse bias, Fig.\ref{fig:gapvsjlzzR}(c). Furthermore, we supplement these panels with Fig.\ref{fig:gapvsjlzzR}(a) in which we plot the rectification $\Rec$ versus $\Delta$.  We observe that for $\Delta>\Delta_R$ (vertical dashed line), in reverse bias there is a significant number of rapidities which tends towards zero. Furthermore, in forward bias the rapidities tend towards zero, however their proximity to zero, and their density, is significantly smaller.

 \begin{figure}[!htb]
            \includegraphics[trim={0 0.5cm 0 0},clip,width=\columnwidth]{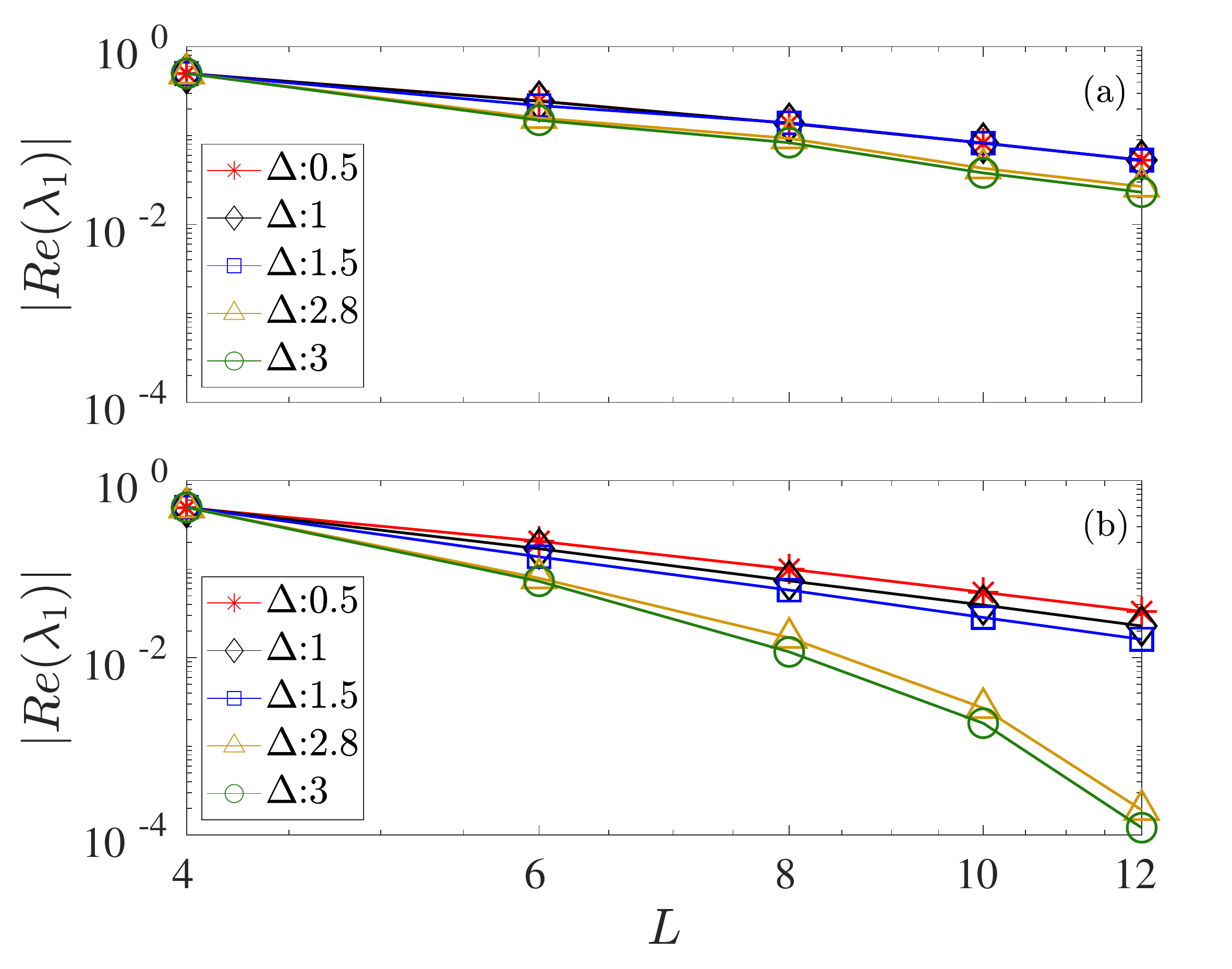}
    \caption{Log-log plot of  $|Re(\lambda_1)|$ against $L$ across $\Delta=0.5,1,1.5,2.8,3$ for forward bias (panel (a)) and reverse bias (panel (b)). Other parameters: $\Gamma=1$, $\gamma=0$, $J_{m}=0.1$.}
    \label{fig:scalinggap}
\end{figure}

To have a more quantitative grasp of how the smallest rapidity approaches zero, in Fig.\ref{fig:scalinggap} we plot the spectral gap versus the system size for different system lengths. In particular, in Fig.\ref{fig:scalinggap}(a) we show the forward bias case, while in Fig.\ref{fig:scalinggap}(b) the reverse bias one. For the smaller value of the anisotropy shown, and the system sizes which we could explore, the decay of the relaxation gap with the system size is consistent with an algebraic decay. For large values of the anisotropy, e.g. $\Delta=2.4$ and $3$, for reverse bias the rapidity seems to decay faster than algebraically, consistent with an expected exponential decay for insulators, while in forward bias the relaxation is in agreement with an algebraic relaxation, although possibly with a different exponent. Though the system sizes we used are not large,  a fairly abrupt change of transport properties are seen in Fig.\ref{fig:loc_magn}(b, d), and the fact that the relaxation gap possibly relaxes exponentially, hint to a possible first order phase transition in the reverse biased system at $\Delta_R$.   

\section{Effect of dephasing}\label{sec:dephasing}

\begin{figure}[!htb]
            \includegraphics[width=\columnwidth]{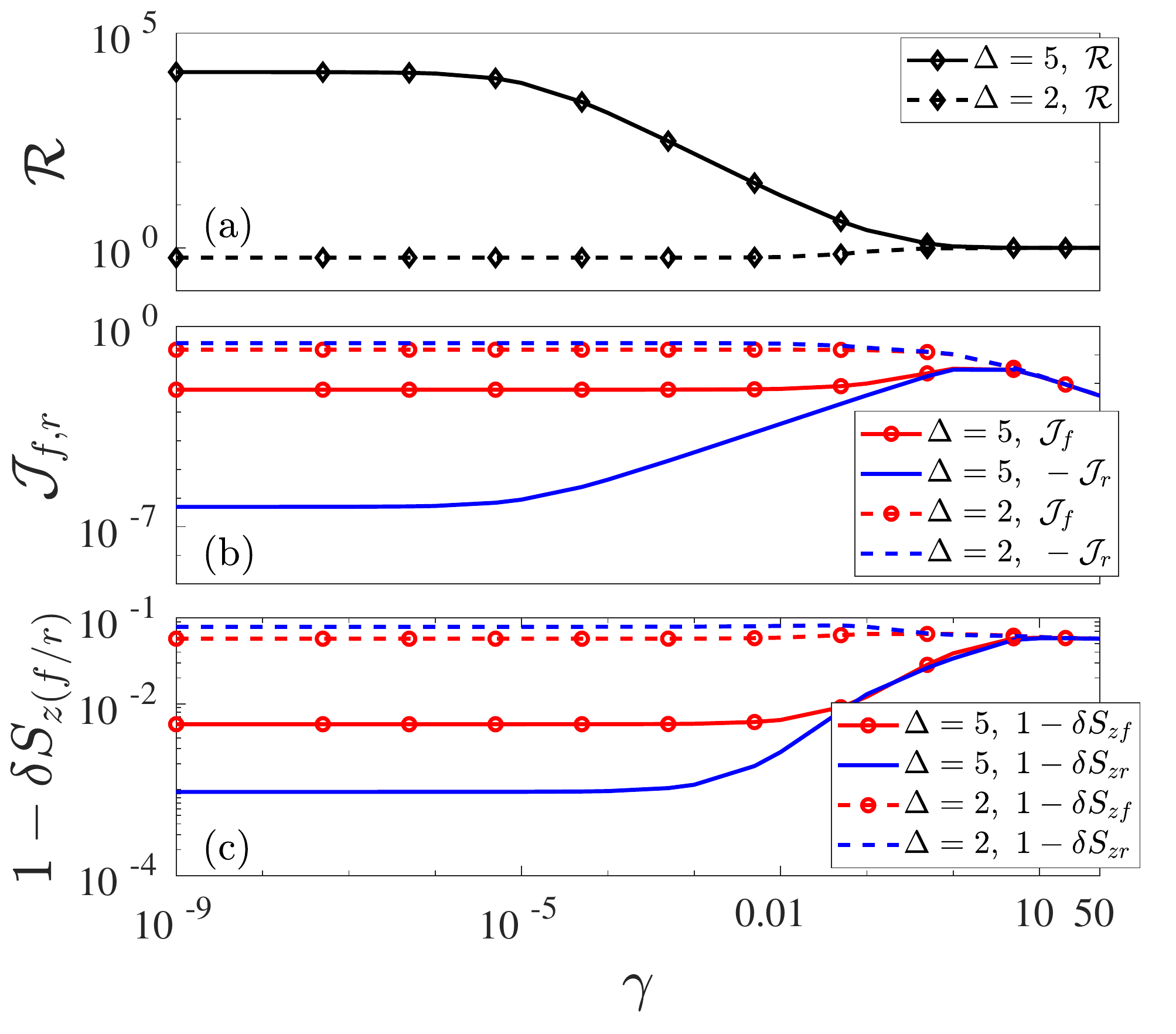}
    \caption{Rectification $\Rec$ (panel (a)), currents $\mathcal{J}_f$ and $\mathcal{J}_r$  (panel (b)) and difference in magnetisation at the interface $\delta S_z$ (panel (c))  as a function of dephasing strength $\gamma$ with $\Delta=5$ (solid) and $\Delta=2$ (dashed).  Forward bias ($\delta S_{zf}$, $\mathcal{J}_f$) are shown by red lines and reverse bias ($\delta S_{zr}$, $\mathcal{J}_r$) with blue lines. Other parameters: $J_m=0.1$, $\Gamma=1$ and $L=8$.}
    \label{fig:rjvsdpjlzz25}
\end{figure} 

Until now we have considered only the boundary driving of $XX+XXZ$ chain in which the baths are trying to impose the system to be in the $\ket{\downarrow}\bra{\downarrow}$ or $(\ket{\downarrow}\bra{\downarrow}+\ket{\uparrow}\bra{\uparrow})/2$. For the case of baths which are not exactly trying to impose these states one can refer to \cite{Balachandran2018}, where it is shown that there is a significant decrease of the rectification. The effect of Hamiltonian perturbations inside the chain have been studied in \cite{lee2021giant}, where it was shown that giant rectification can be obtained even in the presence of Hamiltonian perturbations, however there is no threshold value of anisotropy beyond which the system becomes a perfect rectifier in the thermodynamic limit. For thermal bath it was also shown that significant rectification can emerge, although not to the extent of having a perfect rectifier in the thermodynamic limit \cite{Balachandran2019a}. 

The nature of the dissipation, whether it acts in the bulk or only at the boundaries, may significantly change the transport as well as the spectral properties. It has been shown in a number of articles \cite{Znidaric2010, Mendoza2013, MendozaArenasJaksch2013}, that dephasing (bulk dissipation) leads to diffusive transport, independent of the value of $\Delta$. Furthermore, the scaling of spectral gap can change from exponential to algebraic in a (not-segmented) XXZ chain in the presence of dephasing for $\Delta>1$ \cite{Znidaric2015}. Hence it is relevant to study how dephasing affects the properties that we have previously analyzed. For this, we add a dephasing term on each site as in Eq.(\ref{eq:masterequation}) for $\gamma \ne 0$ in Eq.(\ref{eq:dephasing}).

In Fig.\ref{fig:rjvsdpjlzz25} we consider the rectification $\Rec$, panel (a), the forward (red lines) and reverse (blue lines) bias currents, panel (b), and the magnetization jump at the interface between the XXZ and the XX parts of the chain, panel (c). It is clear from Fig.\ref{fig:rjvsdpjlzz25}(a) that even for values of $\gamma \approx 10^{-3}$ there is a significant reduction of rectification for $\Delta=5>\Delta_R$ whereas  $\Rec$ does not show a strong dependence on $\gamma$ for a small $\Delta=2<\Delta_R$, as there was no significant rectification for these values of the anisotropy. 

These results can be understood from  Fig.\ref{fig:rjvsdpjlzz25}(b) which depicts the forward and reverse bias current separately. We observe that when the current is significantly suppressed by the anisotropy, dephasing has first the effect of increasing the current (allowing transport which was impeded by energy constraint due to the large anisotropy). Then, after reaching a maximum, both currents decrease as dephasing increases because the dephasing destroys all coherences. In Fig.\ref{fig:rjvsdpjlzz25}(c) we study the dependence on $\gamma$ of the magnetization jump at the interface, $\delta S_z=|\hat{\sigma}_{L/2-1}^z-\hat{\sigma}_{L/2}^z|$. Since this value is contained between $0$ and $1$, in Fig.\ref{fig:rjvsdpjlzz25}(c) we plot $1-\delta S_z$. When $\Delta> \Delta_R$ the system is insulating in the reverse bias leading to a step magnetization profile with a maximal jump at the interface. Hence, a minimum value of  $1-\delta S_z$ is expected and is clearly seen in our studies. For small $\Delta$ in the reverse bias as well as for all  $\Delta$'s in the forward bias the system is diffusive and hence a continuous change of magnetization at the interface is seen resulting in a large value of  $1-\delta S_z$. However, we note that finite size effects lead to large jump at the interface for the system size we explore in these cases. Finally, as we increase $\gamma$, $\delta S_z$ becomes identical in forward and reverse bias.

\begin{figure}[!htb]
            \includegraphics[width=\columnwidth]{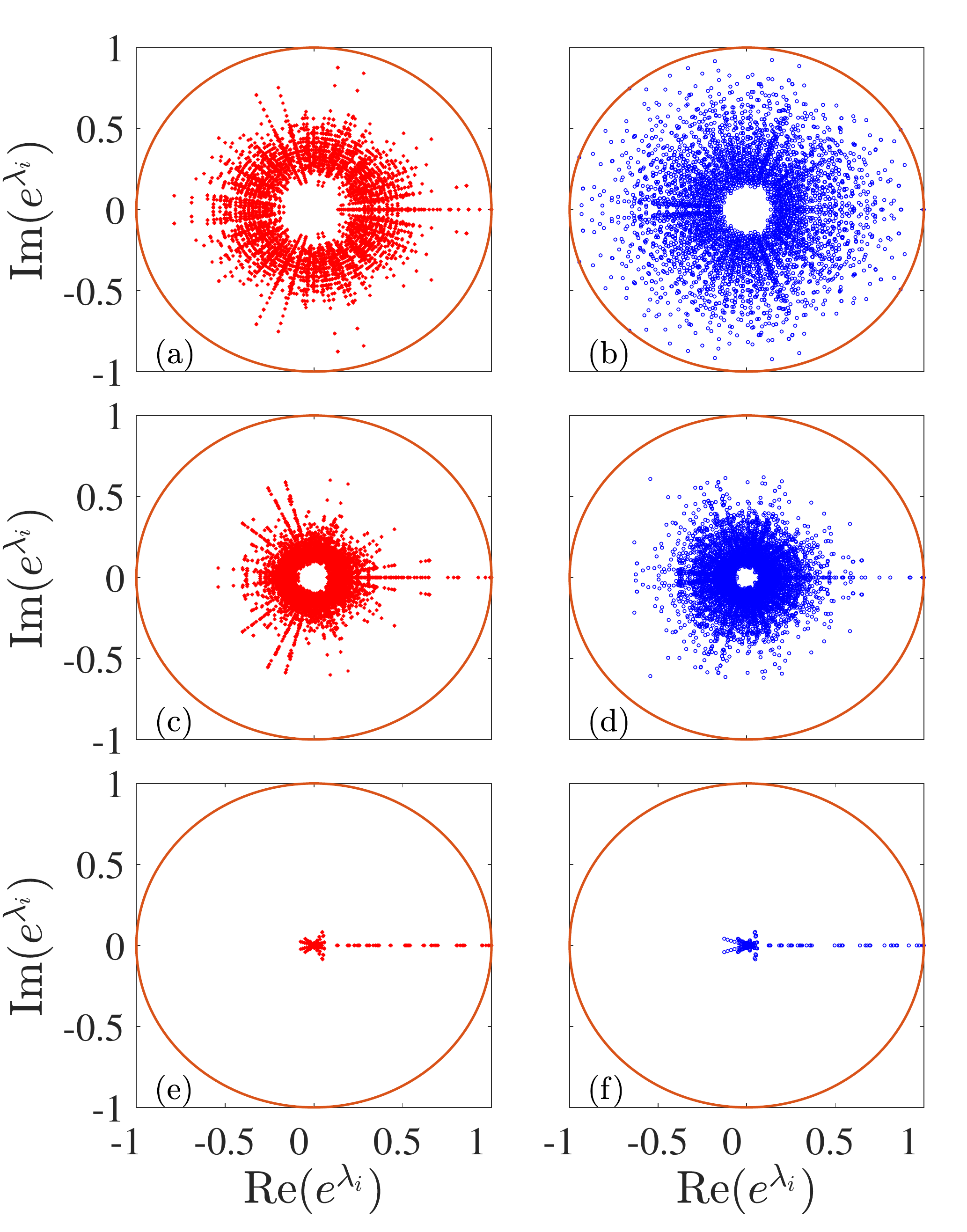}
            \caption{Exponential of rapidities in complex plane for different dephasing rates $\gamma=0$ (a,b), $\gamma=0.1$ (c,d), $\gamma=1$ (e,f). Left panels are for forward bias and right panels for reverse bias. Here, $L=8$, $J_m=0.1$, $\Gamma=1$ and $\Delta=18$.}
            \label{fig:rapidityscatter1}

\end{figure} 

Next, we analyze how dephasing affects the relaxation spectrum. In Fig.\ref{fig:rapidityscatter1} we plot $e^{\lambda_i}$, similarly to Fig.\ref{fig:spectrum}, for forward (left panels) and reverse (right panels) bias. Fig.\ref{fig:rapidityscatter1}(a,b) is for $\gamma=0$ (absence of dephasing), and the lower panels have increasingly larger values of dephasing rate $\gamma$, i.e. $\gamma=0.1$ in Fig.\ref{fig:rapidityscatter1}(c,d) and $\gamma=1$ in Fig.\ref{fig:rapidityscatter1}(e,f). In all panels we use a very large anisotropy $\Delta=18$ to further emphasize the significant impact of the dephasing rate $\gamma$. 
It is very apparent from Fig.\ref{fig:rapidityscatter1} that dephasing forces the rapidities to become real except in the surrounding of the origin, thus suppressing oscillatory dynamics. Dephasing also significantly reduces the number of rapidities with real part close to zero. Importantly, already for $\gamma \approx 1$ there is no apparent difference between the forward and reverse biases, despite $\Delta=18 \gg \Delta_R$. 

\begin{figure}[!htb]
            \includegraphics[width=\columnwidth]{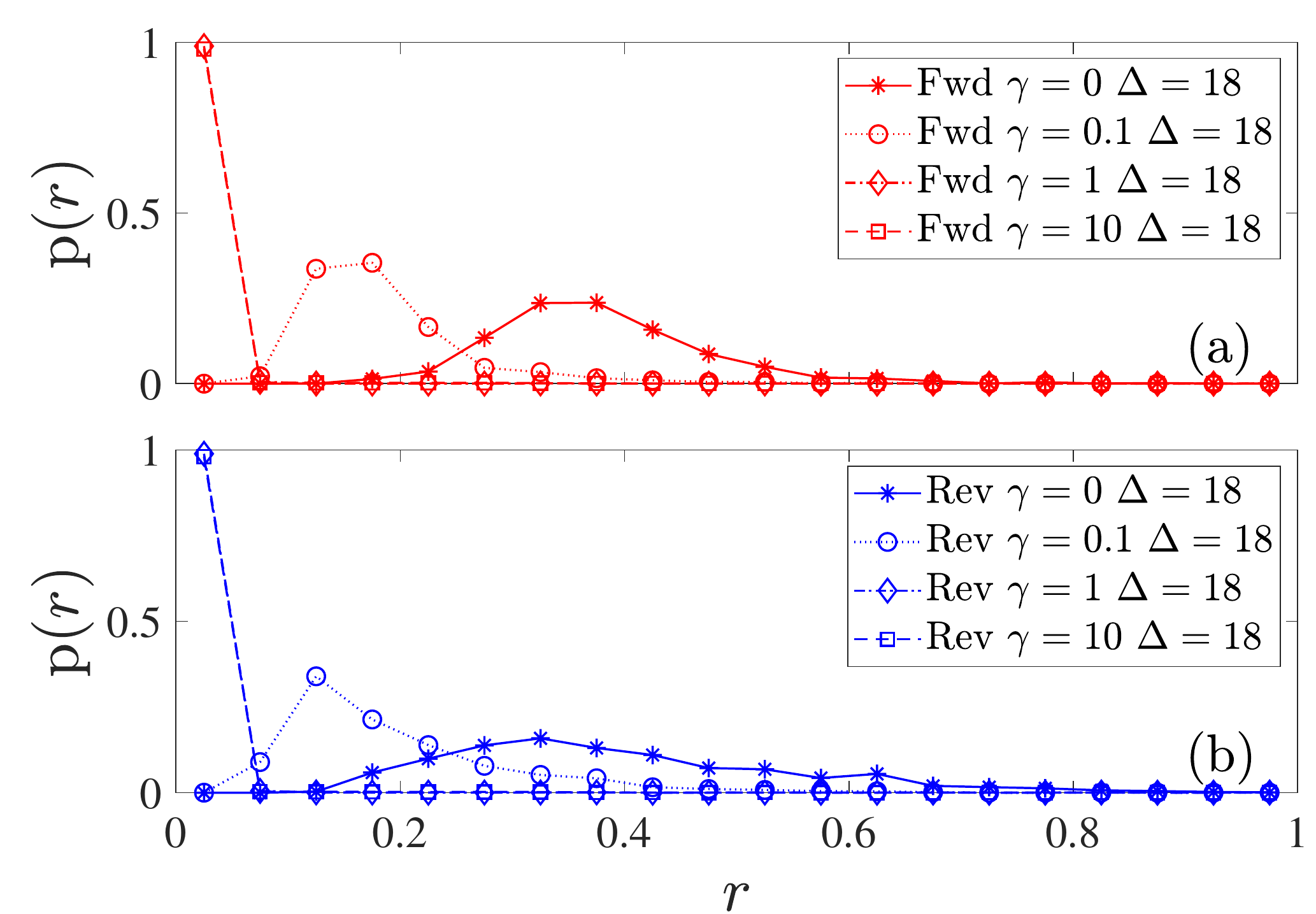}
            \caption{Distribution of rapidities $p(r)$ against the radial direction $r$ in complex plane for $L=8$, $J_m=0.1$, $\Gamma=1$ for different dephasing rates $\gamma$ in the forward direction (a) and reverse direction (b).}
            \label{fig:distrapidities4}

\end{figure} 

To have a better grasp of the distribution of the rapidities, we plot them in Fig.\ref{fig:distrapidities4}. Fig.\ref{fig:distrapidities4}(a) corresponds to the forward bias, while Fig.\ref{fig:distrapidities4}(b) to the reverse bias. We observe that for small dephasing rate the distribution is different in two biases. However, as the $\gamma$ increases the distribution becomes similar and is completely localized in the region around $r=0$. 

\begin{figure}[!htb]
            \includegraphics[width=\columnwidth]{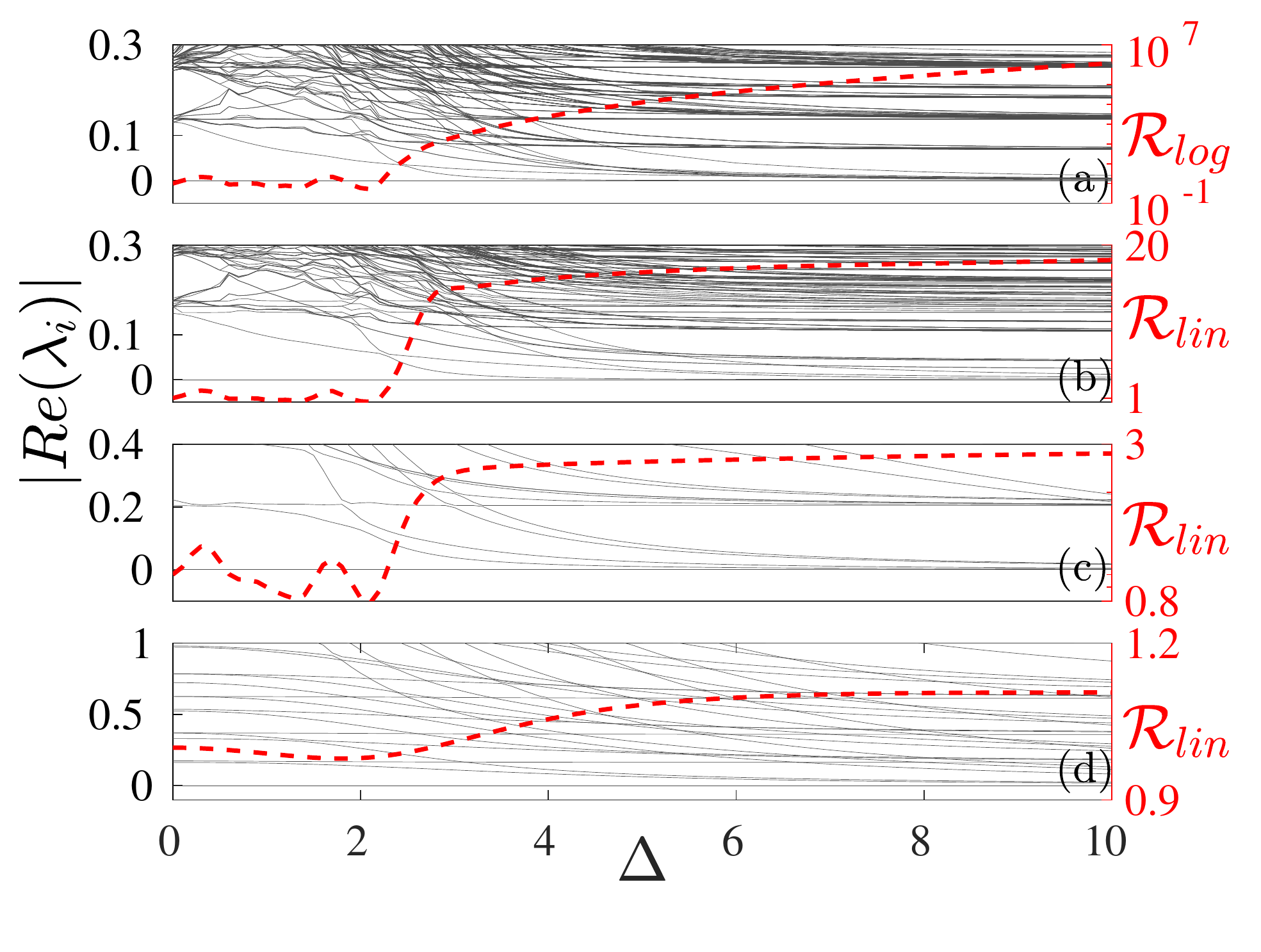}
    \caption{Rapidities  $|Re(\lambda_i)|$ vs $\Delta$ for $\gamma=0$ (a), $\gamma=0.01$ (b),$\gamma=0.1$ (c),$\gamma=1$ (d). Rectification is shown by red dashed lines on the right axis. $\mathcal{R}_{log}$ and $\mathcal{R}_{lin}$ refers to $\mathcal{R}$ plotted in logarithmic and linear scale respectively. Other parameters are $L=8$, $J_m=0.1$ and $\Gamma=1$. }
    \label{fig:gapvsjlzzRdep}
\end{figure} 

We then analyze more in detail the real part of the rapidities $|Re(\lambda_i)|$ versus $\Delta$ for different magnitudes of the dephasing in Fig.\ref{fig:gapvsjlzzRdep}, where we also show the rectification $\Rec$  with a red-dashed line. Each panel corresponds to a different value of the dephasing rate $\gamma$, specifically Fig. \ref{fig:gapvsjlzzRdep}(a-d) correspond to $\gamma = 0,\;0.01,\; 0.1$ and $1$ respectively. Rapidities are plotted only for the reverse bias case as the rectification is significantly determined by reverse bias. From the figure we see that $\Rec$ is significantly affected by dephasing and reaches the value of $\Rec\approx 20$ for $\gamma= 0.01$ and of $\Rec\approx 3$ for $\gamma= 0.1$. From studying the real part of the rapidities, we observe that the density of rapidities close to zero significantly reduces as $\gamma$ increases, indicating faster relaxation and it also corresponds to the disappearance of the insulating regime.   

\begin{figure}[!h]
    \includegraphics[width=\columnwidth]{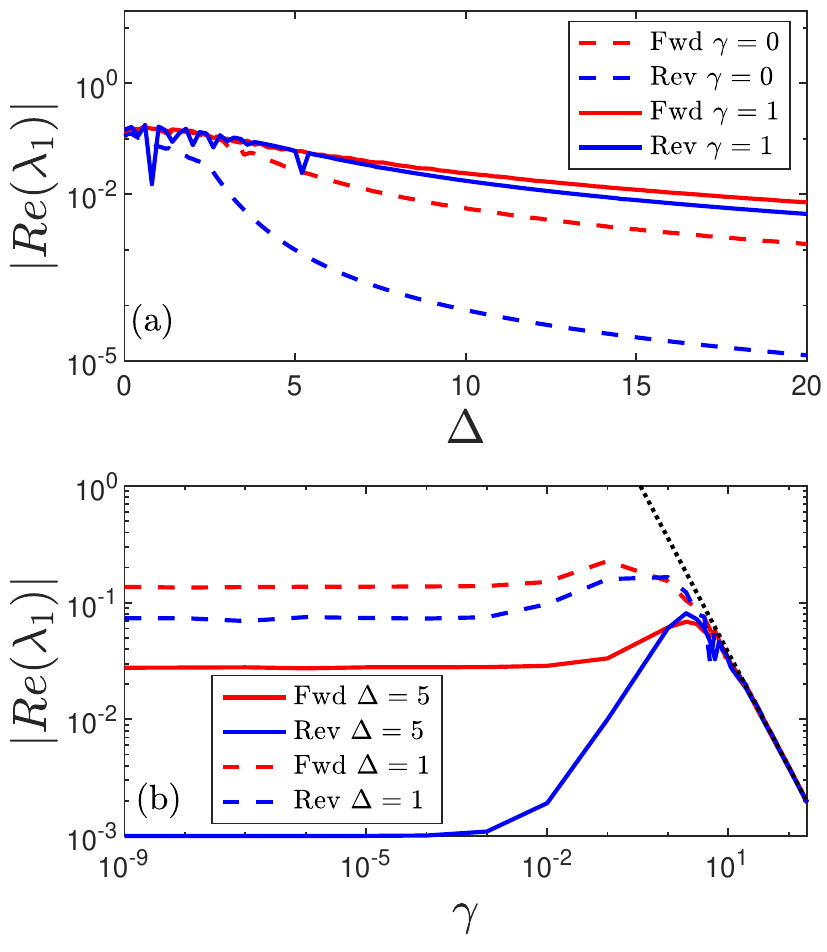}
    \caption{Panel (a): Relaxation Gap $|Re(\lambda_1)|$ against $\Delta$ for different dephasing rates $\gamma=0$ (dashed), $\gamma=1$ (solid). Panel (b): Gap $|Re(\lambda_1)|$ against $\gamma$ for different anisotropies $\Delta=1$ (dashed) and $\Delta=5$ (solid). Other parameters: $L=8$, $J_{m}=0.1$. Black dotted line indicates the inverse scaling of relaxation gap with dephasing rate.}
    \label{fig:gapvsjlzzgamma}
\end{figure}

Last we focus only on the rapidity with the smallest real part different from zero, the relaxation gap. In Fig.\ref{fig:gapvsjlzzgamma}(a,b) we plot the relaxation gap versus $\Delta$ for different values of $\gamma$, panel (a), and versus $\gamma$ for different values of $\Delta$, panel (b). In Fig.\ref{fig:gapvsjlzzgamma}(a) we observe clearly that the relaxation gap increases for larger (but not too large) $\gamma$, and it acquires similar values in the forward (red lines) and reverse biases (blue lines). This is an example of dephasing-assisted tunneling (see for instance \cite{Mendoza-Arenas2013}). 
In Fig.\ref{fig:gapvsjlzzgamma}(b) we also observe how very large values of the dephasing rate $\gamma$ can lead to a smaller relaxation gap which is identical in forward or reverse bias. In particular, the relaxation gap scales as the inverse of the dephasing rate, thus corresponding to quantum Zeno physics \cite{Znidaric2015, syassen2008strong,garcia2009dissipation, PolettiKollath2012}.

\section{Conclusions}\label{sec:conclusions} 

In this work we have studied the transport and spectral properties of the XX+XXZ diode. This diode consisting of a segmented chain coupled to magnetization baths trying to impose the spin down state on one side and an infinite temperature state on the other side has been shown to perform perfectly in the thermodynamic limit. Large rectification stems from the insulating behavior of the system in the reverse bias when anisotropy $\Delta>\Delta_R$. Here we have shown that this setup is even richer. Our analysis of the spin current and local magnetization point towards an understanding that in the forward bias, transport goes from ballistic to diffusive at $\Delta=\Delta_{XXZ}$. For larger anisotropies the spin current decreases significantly, and this is mainly due to an increase in the interface resistance between the two parts. However, because of the system sizes that can be reached, we are not able to conclude whether transport becomes sub-diffusive. At the same time, the analysis of the magnetization is still consistent with a diffusive behavior. 

In reverse bias, the data indicates that there could be three different regimes: ballistic for $\Delta<\Delta_{XXZ}$, diffusive for $\Delta_{XXZ}<\Delta<\Delta_R$ and insulating for $\Delta>\Delta_R$. More studies are required to better characterize the intermediate region with $\Delta_{XXZ}<\Delta<\Delta_R$ as the magnetization profile, for the system sizes we could reach, does not show a clear linear slope as it does in forward bias.  

We also focused our attention on the spectral properties of the system. We have seen that the rapidities tend to different distributions in the forward and reverse bias. In particular, in reverse bias the distributions have far more rapidities with real part close to zero (i.e. the exponential of the rapidity is near the unit circle in the complex plane). This is expected as the system relaxes much more slowly in this regime. In the future it could be interesting to derive an expression for such different distributions and its relation to different transport regimes. 

We then focused on the relaxation gap. Unlike for systems with bulk dissipation, for systems in which dissipation only acts at the boundary, the spectral gap goes to zero with the system size. However, one would expect a change in the dependence of the spectral gap with the system size at a transition, and/or in different phases. We found that in reverse bias the spectral gap goes to zero much faster for $\Delta>\Delta_R$, in a manner consistent with an exponential decay. Furthermore, the density of rapidities close to zero is far more dense in reverse bias compared to forward bias.  

We also investigate the effect of bulk dephasing on the performance of the XX+XXZ diode and its spectral properties. Our analysis shows that dephasing degrades significantly and rapidly the rectifying property of the diode. As for the spectral properties, dephasing results in the exponential of the rapidities to be centered in the origin of the unit circle, independent of whether the system is in forward or reverse bias. Furthermore, we are able to capture the emergence of a quantum Zeno regime in which the spectral gap becomes inversely proportional to the dephasing rate. 

We stress that our insights are captured from small to medium-scale systems and  could serve as a good starting point for more quantitative characterization of the intriguing properties of the system. For instance, a future research direction is how to obtain an accurate characterization of (almost) insulating boundary driven non-equilibrium steady states for which multiple rapidities are extremely close to zero. 
As mentioned in the introduction, XXZ chains can be realized in ultracold atoms experiments. We however note that reaching a large anisotropy may be challenging. For instance in \cite{scholl2021microwave} the largest anisotropy reached is $\Delta=2$. However, another setup in which this physics can be tested is in superconducting circuits, as proposed in \cite{PhysRevLett.126.077203, poulsen2021entanglement}.

\begin{acknowledgments}
D.P. acknowledges support from the Ministry of Education of Singapore AcRF MOE Tier-II (Project No. MOE2016-T2-1-065). 
\end{acknowledgments}

\bibliography{biblio_review}

\begin{thebibliography}{42}%
\makeatletter
\providecommand \@ifxundefined [1]{%
 \@ifx{#1\undefined}
}%
\providecommand \@ifnum [1]{%
 \ifnum #1\expandafter \@firstoftwo
 \else \expandafter \@secondoftwo
 \fi
}%
\providecommand \@ifx [1]{%
 \ifx #1\expandafter \@firstoftwo
 \else \expandafter \@secondoftwo
 \fi
}%
\providecommand \natexlab [1]{#1}%
\providecommand \enquote  [1]{``#1''}%
\providecommand \bibnamefont  [1]{#1}%
\providecommand \bibfnamefont [1]{#1}%
\providecommand \citenamefont [1]{#1}%
\providecommand \href@noop [0]{\@secondoftwo}%
\providecommand \href [0]{\begingroup \@sanitize@url \@href}%
\providecommand \@href[1]{\@@startlink{#1}\@@href}%
\providecommand \@@href[1]{\endgroup#1\@@endlink}%
\providecommand \@sanitize@url [0]{\catcode `\\12\catcode `\$12\catcode
  `\&12\catcode `\#12\catcode `\^12\catcode `\_12\catcode `\%12\relax}%
\providecommand \@@startlink[1]{}%
\providecommand \@@endlink[0]{}%
\providecommand \url  [0]{\begingroup\@sanitize@url \@url }%
\providecommand \@url [1]{\endgroup\@href {#1}{\urlprefix }}%
\providecommand \urlprefix  [0]{URL }%
\providecommand \Eprint [0]{\href }%
\providecommand \doibase [0]{http://dx.doi.org/}%
\providecommand \selectlanguage [0]{\@gobble}%
\providecommand \bibinfo  [0]{\@secondoftwo}%
\providecommand \bibfield  [0]{\@secondoftwo}%
\providecommand \translation [1]{[#1]}%
\providecommand \BibitemOpen [0]{}%
\providecommand \bibitemStop [0]{}%
\providecommand \bibitemNoStop [0]{.\EOS\space}%
\providecommand \EOS [0]{\spacefactor3000\relax}%
\providecommand \BibitemShut  [1]{\csname bibitem#1\endcsname}%
\let\auto@bib@innerbib\@empty
\bibitem [{\citenamefont {Bertini}\ \emph {et~al.}(2020)\citenamefont
  {Bertini}, \citenamefont {Heidrich-Meisner}, \citenamefont {Karrasch},
  \citenamefont {Prosen}, \citenamefont {Steinigeweg},\ and\ \citenamefont
  {Znidaric}}]{BertiniZnidaric2020}%
  \BibitemOpen
  \bibfield  {author} {\bibinfo {author} {\bibfnamefont {B.}~\bibnamefont
  {Bertini}}, \bibinfo {author} {\bibfnamefont {F.}~\bibnamefont
  {Heidrich-Meisner}}, \bibinfo {author} {\bibfnamefont {C.}~\bibnamefont
  {Karrasch}}, \bibinfo {author} {\bibfnamefont {T.}~\bibnamefont {Prosen}},
  \bibinfo {author} {\bibfnamefont {R.}~\bibnamefont {Steinigeweg}}, \ and\
  \bibinfo {author} {\bibfnamefont {M.}~\bibnamefont {Znidaric}},\ }\href
  {arxiv.org/abs/2003.03334} {\bibfield  {journal} {\bibinfo  {journal}
  {arXiv:2003.03334}\ } (\bibinfo {year} {2020})}\BibitemShut {NoStop}%
\bibitem [{\citenamefont {Landi}\ \emph {et~al.}(2021)\citenamefont {Landi},
  \citenamefont {Poletti},\ and\ \citenamefont
  {Schaller}}]{landi2021nonequilibrium}%
  \BibitemOpen
  \bibfield  {author} {\bibinfo {author} {\bibfnamefont {G.~T.}\ \bibnamefont
  {Landi}}, \bibinfo {author} {\bibfnamefont {D.}~\bibnamefont {Poletti}}, \
  and\ \bibinfo {author} {\bibfnamefont {G.}~\bibnamefont {Schaller}},\
  }\href@noop {} {\enquote {\bibinfo {title} {Non-equilibrium boundary driven
  quantum systems: models, methods and properties},}\ } (\bibinfo {year}
  {2021}),\ \Eprint {http://arxiv.org/abs/2104.14350} {arXiv:2104.14350
  [quant-ph]} \BibitemShut {NoStop}%
\bibitem [{\citenamefont {Benenti}\ \emph {et~al.}(2017)\citenamefont
  {Benenti}, \citenamefont {Casati}, \citenamefont {Saito},\ and\ \citenamefont
  {Whitney}}]{BenentiWhitney2017}%
  \BibitemOpen
  \bibfield  {author} {\bibinfo {author} {\bibfnamefont {G.}~\bibnamefont
  {Benenti}}, \bibinfo {author} {\bibfnamefont {G.}~\bibnamefont {Casati}},
  \bibinfo {author} {\bibfnamefont {K.}~\bibnamefont {Saito}}, \ and\ \bibinfo
  {author} {\bibfnamefont {R.~S.}\ \bibnamefont {Whitney}},\ }\href {\doibase
  10.1016/j.physrep.2017.05.008} {\bibfield  {journal} {\bibinfo  {journal}
  {Physics Reports}\ }\textbf {\bibinfo {volume} {694}},\ \bibinfo {pages} {1}
  (\bibinfo {year} {2017})}\BibitemShut {NoStop}%
\bibitem [{\citenamefont {Prosen}(2011{\natexlab{a}})}]{Prosen2011}%
  \BibitemOpen
  \bibfield  {author} {\bibinfo {author} {\bibfnamefont {T.}~\bibnamefont
  {Prosen}},\ }\href {\doibase 10.1103/PhysRevLett.106.217206} {\bibfield
  {journal} {\bibinfo  {journal} {Phys. Rev. Lett.}\ }\textbf {\bibinfo
  {volume} {106}},\ \bibinfo {pages} {217206} (\bibinfo {year}
  {2011}{\natexlab{a}})}\BibitemShut {NoStop}%
\bibitem [{\citenamefont {Benenti}\ \emph {et~al.}(2009)\citenamefont
  {Benenti}, \citenamefont {Casati}, \citenamefont {Prosen},\ and\
  \citenamefont {Rossini}}]{BenentiRossini2009}%
  \BibitemOpen
  \bibfield  {author} {\bibinfo {author} {\bibfnamefont {G.}~\bibnamefont
  {Benenti}}, \bibinfo {author} {\bibfnamefont {G.}~\bibnamefont {Casati}},
  \bibinfo {author} {\bibfnamefont {T.}~\bibnamefont {Prosen}}, \ and\ \bibinfo
  {author} {\bibfnamefont {D.}~\bibnamefont {Rossini}},\ }\href {\doibase
  10.1209/0295-5075/85/37001} {\bibfield  {journal} {\bibinfo  {journal} {{EPL}
  (Europhysics Letters)}\ }\textbf {\bibinfo {volume} {85}},\ \bibinfo {pages}
  {37001} (\bibinfo {year} {2009})}\BibitemShut {NoStop}%
\bibitem [{\citenamefont {Ebadi}\ \emph {et~al.}(2021)\citenamefont {Ebadi},
  \citenamefont {Wang}, \citenamefont {Levine}, \citenamefont {Keesling},
  \citenamefont {Semeghini}, \citenamefont {Omran}, \citenamefont {Bluvstein},
  \citenamefont {Samajdar}, \citenamefont {Pichler}, \citenamefont {Ho} \emph
  {et~al.}}]{ebadi2021quantum}%
  \BibitemOpen
  \bibfield  {author} {\bibinfo {author} {\bibfnamefont {S.}~\bibnamefont
  {Ebadi}}, \bibinfo {author} {\bibfnamefont {T.~T.}\ \bibnamefont {Wang}},
  \bibinfo {author} {\bibfnamefont {H.}~\bibnamefont {Levine}}, \bibinfo
  {author} {\bibfnamefont {A.}~\bibnamefont {Keesling}}, \bibinfo {author}
  {\bibfnamefont {G.}~\bibnamefont {Semeghini}}, \bibinfo {author}
  {\bibfnamefont {A.}~\bibnamefont {Omran}}, \bibinfo {author} {\bibfnamefont
  {D.}~\bibnamefont {Bluvstein}}, \bibinfo {author} {\bibfnamefont
  {R.}~\bibnamefont {Samajdar}}, \bibinfo {author} {\bibfnamefont
  {H.}~\bibnamefont {Pichler}}, \bibinfo {author} {\bibfnamefont {W.~W.}\
  \bibnamefont {Ho}},  \emph {et~al.},\ }\href {\doibase
  10.1038/s41586-021-03582-4} {\bibfield  {journal} {\bibinfo  {journal}
  {Nature}\ }\textbf {\bibinfo {volume} {595}},\ \bibinfo {pages} {227}
  (\bibinfo {year} {2021})}\BibitemShut {NoStop}%
\bibitem [{\citenamefont {Scholl}\ \emph {et~al.}(2021)\citenamefont {Scholl},
  \citenamefont {Williams}, \citenamefont {Bornet}, \citenamefont {Wallner},
  \citenamefont {Barredo}, \citenamefont {Lahaye}, \citenamefont {Browaeys},
  \citenamefont {Henriet}, \citenamefont {Signoles}, \citenamefont {Hainaut}
  \emph {et~al.}}]{scholl2021microwave}%
  \BibitemOpen
  \bibfield  {author} {\bibinfo {author} {\bibfnamefont {P.}~\bibnamefont
  {Scholl}}, \bibinfo {author} {\bibfnamefont {H.}~\bibnamefont {Williams}},
  \bibinfo {author} {\bibfnamefont {G.}~\bibnamefont {Bornet}}, \bibinfo
  {author} {\bibfnamefont {F.}~\bibnamefont {Wallner}}, \bibinfo {author}
  {\bibfnamefont {D.}~\bibnamefont {Barredo}}, \bibinfo {author} {\bibfnamefont
  {T.}~\bibnamefont {Lahaye}}, \bibinfo {author} {\bibfnamefont
  {A.}~\bibnamefont {Browaeys}}, \bibinfo {author} {\bibfnamefont
  {L.}~\bibnamefont {Henriet}}, \bibinfo {author} {\bibfnamefont
  {A.}~\bibnamefont {Signoles}}, \bibinfo {author} {\bibfnamefont
  {C.}~\bibnamefont {Hainaut}},  \emph {et~al.},\ }\href@noop {} {\bibfield
  {journal} {\bibinfo  {journal} {arXiv preprint arXiv:2107.14459}\ } (\bibinfo
  {year} {2021})}\BibitemShut {NoStop}%
\bibitem [{\citenamefont {Jepsen}\ \emph {et~al.}(2020)\citenamefont {Jepsen},
  \citenamefont {Amato-Grill}, \citenamefont {Dimitrova}, \citenamefont {Ho},
  \citenamefont {Demler},\ and\ \citenamefont {Ketterle}}]{Jepsennature2020}%
  \BibitemOpen
  \bibfield  {author} {\bibinfo {author} {\bibfnamefont {P.~N.}\ \bibnamefont
  {Jepsen}}, \bibinfo {author} {\bibfnamefont {J.}~\bibnamefont {Amato-Grill}},
  \bibinfo {author} {\bibfnamefont {I.}~\bibnamefont {Dimitrova}}, \bibinfo
  {author} {\bibfnamefont {W.~W.}\ \bibnamefont {Ho}}, \bibinfo {author}
  {\bibfnamefont {E.}~\bibnamefont {Demler}}, \ and\ \bibinfo {author}
  {\bibfnamefont {W.}~\bibnamefont {Ketterle}},\ }\href {\doibase
  10.1038/s41586-020-3033-y} {\bibfield  {journal} {\bibinfo  {journal}
  {Nature}\ }\textbf {\bibinfo {volume} {588}},\ \bibinfo {pages} {403–407}
  (\bibinfo {year} {2020})}\BibitemShut {NoStop}%
\bibitem [{\citenamefont {Poulsen}\ and\ \citenamefont
  {Zinner}(2021)}]{PhysRevLett.126.077203}%
  \BibitemOpen
  \bibfield  {author} {\bibinfo {author} {\bibfnamefont {K.}~\bibnamefont
  {Poulsen}}\ and\ \bibinfo {author} {\bibfnamefont {N.~T.}\ \bibnamefont
  {Zinner}},\ }\href {\doibase 10.1103/PhysRevLett.126.077203} {\bibfield
  {journal} {\bibinfo  {journal} {Phys. Rev. Lett.}\ }\textbf {\bibinfo
  {volume} {126}},\ \bibinfo {pages} {077203} (\bibinfo {year}
  {2021})}\BibitemShut {NoStop}%
\bibitem [{\citenamefont {Poulsen}\ \emph {et~al.}(2021)\citenamefont
  {Poulsen}, \citenamefont {Santos}, \citenamefont {Kristensen},\ and\
  \citenamefont {Zinner}}]{poulsen2021entanglement}%
  \BibitemOpen
  \bibfield  {author} {\bibinfo {author} {\bibfnamefont {K.}~\bibnamefont
  {Poulsen}}, \bibinfo {author} {\bibfnamefont {A.~C.}\ \bibnamefont {Santos}},
  \bibinfo {author} {\bibfnamefont {L.~B.}\ \bibnamefont {Kristensen}}, \ and\
  \bibinfo {author} {\bibfnamefont {N.~T.}\ \bibnamefont {Zinner}},\
  }\href@noop {} {\bibfield  {journal} {\bibinfo  {journal} {arXiv preprint
  arXiv:2101.04124}\ } (\bibinfo {year} {2021})}\BibitemShut {NoStop}%
\bibitem [{\citenamefont {Balachandran}\ \emph {et~al.}(2018)\citenamefont
  {Balachandran}, \citenamefont {Benenti}, \citenamefont {Pereira},
  \citenamefont {Casati},\ and\ \citenamefont {Poletti}}]{Balachandran2018}%
  \BibitemOpen
  \bibfield  {author} {\bibinfo {author} {\bibfnamefont {V.}~\bibnamefont
  {Balachandran}}, \bibinfo {author} {\bibfnamefont {G.}~\bibnamefont
  {Benenti}}, \bibinfo {author} {\bibfnamefont {E.}~\bibnamefont {Pereira}},
  \bibinfo {author} {\bibfnamefont {G.}~\bibnamefont {Casati}}, \ and\ \bibinfo
  {author} {\bibfnamefont {D.}~\bibnamefont {Poletti}},\ }\href {\doibase
  10.1103/PhysRevLett.120.200603} {\bibfield  {journal} {\bibinfo  {journal}
  {Physical Review Letters}\ }\textbf {\bibinfo {volume} {120}},\ \bibinfo
  {pages} {200603} (\bibinfo {year} {2018})}\BibitemShut {NoStop}%
\bibitem [{\citenamefont {Balachandran}\ \emph
  {et~al.}(2019{\natexlab{a}})\citenamefont {Balachandran}, \citenamefont
  {Clark}, \citenamefont {Goold},\ and\ \citenamefont
  {Poletti}}]{Balachandran2019}%
  \BibitemOpen
  \bibfield  {author} {\bibinfo {author} {\bibfnamefont {V.}~\bibnamefont
  {Balachandran}}, \bibinfo {author} {\bibfnamefont {S.~R.}\ \bibnamefont
  {Clark}}, \bibinfo {author} {\bibfnamefont {J.}~\bibnamefont {Goold}}, \ and\
  \bibinfo {author} {\bibfnamefont {D.}~\bibnamefont {Poletti}},\ }\href
  {\doibase 10.1103/PhysRevLett.123.020603} {\bibfield  {journal} {\bibinfo
  {journal} {Physical Review Letters}\ }\textbf {\bibinfo {volume} {123}},\
  \bibinfo {pages} {020603} (\bibinfo {year} {2019}{\natexlab{a}})}\BibitemShut
  {NoStop}%
\bibitem [{\citenamefont {Lee}\ \emph {et~al.}(2020)\citenamefont {Lee},
  \citenamefont {Balachandran}, \citenamefont {Guo},\ and\ \citenamefont
  {Poletti}}]{LeePoletti2019}%
  \BibitemOpen
  \bibfield  {author} {\bibinfo {author} {\bibfnamefont {K.}~\bibnamefont
  {Lee}}, \bibinfo {author} {\bibfnamefont {V.}~\bibnamefont {Balachandran}},
  \bibinfo {author} {\bibfnamefont {C.}~\bibnamefont {Guo}}, \ and\ \bibinfo
  {author} {\bibfnamefont {D.}~\bibnamefont {Poletti}},\ }\href {\doibase
  10.3390/e22111311} {\bibfield  {journal} {\bibinfo  {journal} {Entropy}\
  }\textbf {\bibinfo {volume} {22}},\ \bibinfo {pages} {1311} (\bibinfo {year}
  {2020})}\BibitemShut {NoStop}%
\bibitem [{\citenamefont {Lee}\ \emph {et~al.}(2021)\citenamefont {Lee},
  \citenamefont {Balachandran},\ and\ \citenamefont {Poletti}}]{lee2021giant}%
  \BibitemOpen
  \bibfield  {author} {\bibinfo {author} {\bibfnamefont {K.~H.}\ \bibnamefont
  {Lee}}, \bibinfo {author} {\bibfnamefont {V.}~\bibnamefont {Balachandran}}, \
  and\ \bibinfo {author} {\bibfnamefont {D.}~\bibnamefont {Poletti}},\ }\href
  {\doibase 10.1103/PhysRevE.103.052143} {\bibfield  {journal} {\bibinfo
  {journal} {Physical Review E}\ }\textbf {\bibinfo {volume} {103}},\ \bibinfo
  {pages} {052143} (\bibinfo {year} {2021})}\BibitemShut {NoStop}%
\bibitem [{\citenamefont {Biella}\ \emph {et~al.}(2016)\citenamefont {Biella},
  \citenamefont {De~Luca}, \citenamefont {Viti}, \citenamefont {Rossini},
  \citenamefont {Mazza},\ and\ \citenamefont {Fazio}}]{BiellaFazio2016}%
  \BibitemOpen
  \bibfield  {author} {\bibinfo {author} {\bibfnamefont {A.}~\bibnamefont
  {Biella}}, \bibinfo {author} {\bibfnamefont {A.}~\bibnamefont {De~Luca}},
  \bibinfo {author} {\bibfnamefont {J.}~\bibnamefont {Viti}}, \bibinfo {author}
  {\bibfnamefont {D.}~\bibnamefont {Rossini}}, \bibinfo {author} {\bibfnamefont
  {L.}~\bibnamefont {Mazza}}, \ and\ \bibinfo {author} {\bibfnamefont
  {R.}~\bibnamefont {Fazio}},\ }\href {\doibase 10.1103/PhysRevB.93.205121}
  {\bibfield  {journal} {\bibinfo  {journal} {Phys. Rev. B}\ }\textbf {\bibinfo
  {volume} {93}},\ \bibinfo {pages} {205121} (\bibinfo {year}
  {2016})}\BibitemShut {NoStop}%
\bibitem [{\citenamefont {Biella}\ \emph {et~al.}(2019)\citenamefont {Biella},
  \citenamefont {Collura}, \citenamefont {Rossini}, \citenamefont {De~Luca},\
  and\ \citenamefont {Mazza}}]{BiellaMazza2019}%
  \BibitemOpen
  \bibfield  {author} {\bibinfo {author} {\bibfnamefont {A.}~\bibnamefont
  {Biella}}, \bibinfo {author} {\bibfnamefont {M.}~\bibnamefont {Collura}},
  \bibinfo {author} {\bibfnamefont {D.}~\bibnamefont {Rossini}}, \bibinfo
  {author} {\bibfnamefont {A.}~\bibnamefont {De~Luca}}, \ and\ \bibinfo
  {author} {\bibfnamefont {L.}~\bibnamefont {Mazza}},\ }\href {\doibase
  10.1038/s41467-019-12784-4} {\bibfield  {journal} {\bibinfo  {journal}
  {Nature Communications}\ }\textbf {\bibinfo {volume} {10}},\ \bibinfo {pages}
  {4820} (\bibinfo {year} {2019})}\BibitemShut {NoStop}%
\bibitem [{\citenamefont {Kessler}\ \emph {et~al.}(2012)\citenamefont
  {Kessler}, \citenamefont {Giedke}, \citenamefont {Imamoglu}, \citenamefont
  {Yelin}, \citenamefont {Lukin},\ and\ \citenamefont {Cirac}}]{Kessler2012}%
  \BibitemOpen
  \bibfield  {author} {\bibinfo {author} {\bibfnamefont {E.~M.}\ \bibnamefont
  {Kessler}}, \bibinfo {author} {\bibfnamefont {G.}~\bibnamefont {Giedke}},
  \bibinfo {author} {\bibfnamefont {A.}~\bibnamefont {Imamoglu}}, \bibinfo
  {author} {\bibfnamefont {S.~F.}\ \bibnamefont {Yelin}}, \bibinfo {author}
  {\bibfnamefont {M.~D.}\ \bibnamefont {Lukin}}, \ and\ \bibinfo {author}
  {\bibfnamefont {J.~I.}\ \bibnamefont {Cirac}},\ }\href {\doibase
  10.1103/PhysRevA.86.012116} {\bibfield  {journal} {\bibinfo  {journal} {Phys.
  Rev. A}\ }\textbf {\bibinfo {volume} {86}},\ \bibinfo {pages} {012116}
  (\bibinfo {year} {2012})}\BibitemShut {NoStop}%
\bibitem [{\citenamefont {Minganti}\ \emph {et~al.}(2018)\citenamefont
  {Minganti}, \citenamefont {Biella}, \citenamefont {Bartolo},\ and\
  \citenamefont {Ciuti}}]{Minganti2018}%
  \BibitemOpen
  \bibfield  {author} {\bibinfo {author} {\bibfnamefont {F.}~\bibnamefont
  {Minganti}}, \bibinfo {author} {\bibfnamefont {A.}~\bibnamefont {Biella}},
  \bibinfo {author} {\bibfnamefont {N.}~\bibnamefont {Bartolo}}, \ and\
  \bibinfo {author} {\bibfnamefont {C.}~\bibnamefont {Ciuti}},\ }\href
  {\doibase 10.1103/PhysRevA.98.042118} {\bibfield  {journal} {\bibinfo
  {journal} {Phys. Rev. A}\ }\textbf {\bibinfo {volume} {98}},\ \bibinfo
  {pages} {042118} (\bibinfo {year} {2018})}\BibitemShut {NoStop}%
\bibitem [{\citenamefont {Morrison}\ and\ \citenamefont
  {Parkins}(2008)}]{morrison2008a}%
  \BibitemOpen
  \bibfield  {author} {\bibinfo {author} {\bibfnamefont {S.}~\bibnamefont
  {Morrison}}\ and\ \bibinfo {author} {\bibfnamefont {A.~S.}\ \bibnamefont
  {Parkins}},\ }\href {\doibase 10.1103/PhysRevLett.100.040403} {\bibfield
  {journal} {\bibinfo  {journal} {Physical Review Letters}\ }\textbf {\bibinfo
  {volume} {100}},\ \bibinfo {pages} {040403} (\bibinfo {year}
  {2008})}\BibitemShut {NoStop}%
\bibitem [{\citenamefont {Fitzpatrick}\ \emph {et~al.}(2017)\citenamefont
  {Fitzpatrick}, \citenamefont {Sundaresan}, \citenamefont {Li}, \citenamefont
  {Koch},\ and\ \citenamefont {Houck}}]{FitzpatrickHouck2017}%
  \BibitemOpen
  \bibfield  {author} {\bibinfo {author} {\bibfnamefont {M.}~\bibnamefont
  {Fitzpatrick}}, \bibinfo {author} {\bibfnamefont {N.~M.}\ \bibnamefont
  {Sundaresan}}, \bibinfo {author} {\bibfnamefont {A.~C.~Y.}\ \bibnamefont
  {Li}}, \bibinfo {author} {\bibfnamefont {J.}~\bibnamefont {Koch}}, \ and\
  \bibinfo {author} {\bibfnamefont {A.~A.}\ \bibnamefont {Houck}},\ }\href
  {\doibase 10.1103/PhysRevX.7.011016} {\bibfield  {journal} {\bibinfo
  {journal} {Phys. Rev. X}\ }\textbf {\bibinfo {volume} {7}},\ \bibinfo {pages}
  {011016} (\bibinfo {year} {2017})}\BibitemShut {NoStop}%
\bibitem [{\citenamefont {Lindblad}(1976)}]{Lindblad1976}%
  \BibitemOpen
  \bibfield  {author} {\bibinfo {author} {\bibfnamefont {G.}~\bibnamefont
  {Lindblad}},\ }\href {http://link.springer.com/article/10.1007/BF01608499}
  {\bibfield  {journal} {\bibinfo  {journal} {Communications in Mathematical
  Physics}\ }\textbf {\bibinfo {volume} {48}},\ \bibinfo {pages} {119}
  (\bibinfo {year} {1976})}\BibitemShut {NoStop}%
\bibitem [{\citenamefont {Gorini}\ \emph {et~al.}(1976)\citenamefont {Gorini},
  \citenamefont {Kossakowski},\ and\ \citenamefont {Sudarshan}}]{Gorini1976}%
  \BibitemOpen
  \bibfield  {author} {\bibinfo {author} {\bibfnamefont {V.}~\bibnamefont
  {Gorini}}, \bibinfo {author} {\bibfnamefont {A.}~\bibnamefont {Kossakowski}},
  \ and\ \bibinfo {author} {\bibfnamefont {E.~C.~G.}\ \bibnamefont
  {Sudarshan}},\ }\href {\doibase 10.1063/1.522979} {\bibfield  {journal}
  {\bibinfo  {journal} {Journal of Mathematical Physics}\ }\textbf {\bibinfo
  {volume} {17}},\ \bibinfo {pages} {821} (\bibinfo {year} {1976})}\BibitemShut
  {NoStop}%
\bibitem [{\citenamefont {{\v{Z}nidari\v{c}}}(2015)}]{Znidaric2015}%
  \BibitemOpen
  \bibfield  {author} {\bibinfo {author} {\bibfnamefont {M.}~\bibnamefont
  {{\v{Z}nidari\v{c}}}},\ }\href {\doibase 10.1103/PhysRevE.92.042143}
  {\bibfield  {journal} {\bibinfo  {journal} {Phys. Rev. E}\ }\textbf {\bibinfo
  {volume} {92}},\ \bibinfo {pages} {042143} (\bibinfo {year}
  {2015})}\BibitemShut {NoStop}%
\bibitem [{\citenamefont {White}(1992)}]{White1992}%
  \BibitemOpen
  \bibfield  {author} {\bibinfo {author} {\bibfnamefont {S.~R.}\ \bibnamefont
  {White}},\ }\href {\doibase 10.1103/PhysRevLett.69.2863} {\bibfield
  {journal} {\bibinfo  {journal} {Phys. Rev. Lett.}\ }\textbf {\bibinfo
  {volume} {69}},\ \bibinfo {pages} {2863} (\bibinfo {year}
  {1992})}\BibitemShut {NoStop}%
\bibitem [{\citenamefont {Schollw\"{o}ck}(2011)}]{Schollwock2011}%
  \BibitemOpen
  \bibfield  {author} {\bibinfo {author} {\bibfnamefont {U.}~\bibnamefont
  {Schollw\"{o}ck}},\ }\href {\doibase 10.1016/j.aop.2010.09.012} {\bibfield
  {journal} {\bibinfo  {journal} {Annals of Phys.}\ }\textbf {\bibinfo {volume}
  {326}},\ \bibinfo {pages} {96} (\bibinfo {year} {2011})}\BibitemShut
  {NoStop}%
\bibitem [{foo()}]{footnote1}%
  \BibitemOpen
  \href@noop {} {}\bibinfo {note} {Leveraging on symmetries in the system, see
  \cite{GuoPoletti2017b, SaProsen2020, Znidaric2015}, the scaling is less than
  $2^{2L}$, but at most $[(2L)!/(L!)^2]$. This allows us to study slightly
  larger system sizes}\BibitemShut {NoStop}%
\bibitem [{\citenamefont {Prosen}\ and\ \citenamefont
  {Ilievski}(2013)}]{ProsenIlievski2013}%
  \BibitemOpen
  \bibfield  {author} {\bibinfo {author} {\bibfnamefont {T.}~\bibnamefont
  {Prosen}}\ and\ \bibinfo {author} {\bibfnamefont {E.}~\bibnamefont
  {Ilievski}},\ }\href {\doibase 10.1103/PhysRevLett.111.057203} {\bibfield
  {journal} {\bibinfo  {journal} {Phys. Rev. Lett.}\ }\textbf {\bibinfo
  {volume} {111}},\ \bibinfo {pages} {057203} (\bibinfo {year}
  {2013})}\BibitemShut {NoStop}%
\bibitem [{\citenamefont {Prosen}(2011{\natexlab{b}})}]{Prosen2011b}%
  \BibitemOpen
  \bibfield  {author} {\bibinfo {author} {\bibfnamefont {T.}~\bibnamefont
  {Prosen}},\ }\href {\doibase 10.1103/PhysRevLett.107.137201} {\bibfield
  {journal} {\bibinfo  {journal} {Phys. Rev. Lett.}\ }\textbf {\bibinfo
  {volume} {107}},\ \bibinfo {pages} {137201} (\bibinfo {year}
  {2011}{\natexlab{b}})}\BibitemShut {NoStop}%
\bibitem [{\citenamefont {Prosen}\ and\ \citenamefont
  {\v{Z}nidari\v{c}}(2012)}]{ProsenZnidaric2012}%
  \BibitemOpen
  \bibfield  {author} {\bibinfo {author} {\bibfnamefont {T.}~\bibnamefont
  {Prosen}}\ and\ \bibinfo {author} {\bibfnamefont {M.}~\bibnamefont
  {\v{Z}nidari\v{c}}},\ }\href {\doibase 10.1103/PhysRevB.86.125118} {\bibfield
   {journal} {\bibinfo  {journal} {Phys. Rev. B}\ }\textbf {\bibinfo {volume}
  {86}},\ \bibinfo {pages} {125118} (\bibinfo {year} {2012})}\BibitemShut
  {NoStop}%
\bibitem [{Note1()}]{Note1}%
  \BibitemOpen
  \bibinfo {note} {The maximum absolute value of the real part of the rapidity
  depends upon the number of the sites in the chain to which the baths are
  coupled. Hence in our setup the exponential of the rapidities fall within two
  concentric circles of radii $e^{-2}$ and $1$}\BibitemShut {NoStop}%
\bibitem [{\citenamefont {Casteels}\ \emph {et~al.}(2017)\citenamefont
  {Casteels}, \citenamefont {Fazio},\ and\ \citenamefont
  {Ciuti}}]{Casteels2017}%
  \BibitemOpen
  \bibfield  {author} {\bibinfo {author} {\bibfnamefont {W.}~\bibnamefont
  {Casteels}}, \bibinfo {author} {\bibfnamefont {R.}~\bibnamefont {Fazio}}, \
  and\ \bibinfo {author} {\bibfnamefont {C.}~\bibnamefont {Ciuti}},\ }\href
  {\doibase 10.1103/PhysRevA.95.012128} {\bibfield  {journal} {\bibinfo
  {journal} {Phys. Rev. A}\ }\textbf {\bibinfo {volume} {95}},\ \bibinfo
  {pages} {012128} (\bibinfo {year} {2017})}\BibitemShut {NoStop}%
\bibitem [{\citenamefont {Horstmann}\ \emph {et~al.}(2013)\citenamefont
  {Horstmann}, \citenamefont {Cirac},\ and\ \citenamefont
  {Giedke}}]{Horstmann2013}%
  \BibitemOpen
  \bibfield  {author} {\bibinfo {author} {\bibfnamefont {B.}~\bibnamefont
  {Horstmann}}, \bibinfo {author} {\bibfnamefont {J.~I.}\ \bibnamefont
  {Cirac}}, \ and\ \bibinfo {author} {\bibfnamefont {G.}~\bibnamefont
  {Giedke}},\ }\href {\doibase 10.1103/PhysRevA.87.012108} {\bibfield
  {journal} {\bibinfo  {journal} {Phys. Rev. A}\ }\textbf {\bibinfo {volume}
  {87}},\ \bibinfo {pages} {012108} (\bibinfo {year} {2013})}\BibitemShut
  {NoStop}%
\bibitem [{\citenamefont {Balachandran}\ \emph
  {et~al.}(2019{\natexlab{b}})\citenamefont {Balachandran}, \citenamefont
  {Benenti}, \citenamefont {Pereira}, \citenamefont {Casati},\ and\
  \citenamefont {Poletti}}]{Balachandran2019a}%
  \BibitemOpen
  \bibfield  {author} {\bibinfo {author} {\bibfnamefont {V.}~\bibnamefont
  {Balachandran}}, \bibinfo {author} {\bibfnamefont {G.}~\bibnamefont
  {Benenti}}, \bibinfo {author} {\bibfnamefont {E.}~\bibnamefont {Pereira}},
  \bibinfo {author} {\bibfnamefont {G.}~\bibnamefont {Casati}}, \ and\ \bibinfo
  {author} {\bibfnamefont {D.}~\bibnamefont {Poletti}},\ }\href {\doibase
  10.1103/PhysRevE.99.032136} {\bibfield  {journal} {\bibinfo  {journal}
  {Physical Review E}\ }\textbf {\bibinfo {volume} {99}},\ \bibinfo {pages}
  {032136} (\bibinfo {year} {2019}{\natexlab{b}})}\BibitemShut {NoStop}%
\bibitem [{\citenamefont {{\v{Z}}nidari{\v{c}}}(2010)}]{Znidaric2010}%
  \BibitemOpen
  \bibfield  {author} {\bibinfo {author} {\bibfnamefont {M.}~\bibnamefont
  {{\v{Z}}nidari{\v{c}}}},\ }\href {\doibase 10.1088/1367-2630/12/4/043001}
  {\bibfield  {journal} {\bibinfo  {journal} {New Journal of Physics}\ }\textbf
  {\bibinfo {volume} {12}},\ \bibinfo {pages} {043001} (\bibinfo {year}
  {2010})}\BibitemShut {NoStop}%
\bibitem [{\citenamefont {Mendoza-Arenas}\ \emph
  {et~al.}(2013{\natexlab{a}})\citenamefont {Mendoza-Arenas}, \citenamefont
  {Grujic}, \citenamefont {Jaksch},\ and\ \citenamefont {Clark}}]{Mendoza2013}%
  \BibitemOpen
  \bibfield  {author} {\bibinfo {author} {\bibfnamefont {J.~J.}\ \bibnamefont
  {Mendoza-Arenas}}, \bibinfo {author} {\bibfnamefont {T.}~\bibnamefont
  {Grujic}}, \bibinfo {author} {\bibfnamefont {D.}~\bibnamefont {Jaksch}}, \
  and\ \bibinfo {author} {\bibfnamefont {S.~R.}\ \bibnamefont {Clark}},\ }\href
  {\doibase 10.1103/PhysRevB.87.235130} {\bibfield  {journal} {\bibinfo
  {journal} {Phys. Rev. B}\ }\textbf {\bibinfo {volume} {87}},\ \bibinfo
  {pages} {235130} (\bibinfo {year} {2013}{\natexlab{a}})}\BibitemShut
  {NoStop}%
\bibitem [{\citenamefont {Mendoza-Arenas}\ \emph
  {et~al.}(2013{\natexlab{b}})\citenamefont {Mendoza-Arenas}, \citenamefont
  {Al-Assam}, \citenamefont {Clark},\ and\ \citenamefont
  {Jaksch}}]{MendozaArenasJaksch2013}%
  \BibitemOpen
  \bibfield  {author} {\bibinfo {author} {\bibfnamefont {J.~J.}\ \bibnamefont
  {Mendoza-Arenas}}, \bibinfo {author} {\bibfnamefont {S.}~\bibnamefont
  {Al-Assam}}, \bibinfo {author} {\bibfnamefont {S.~R.}\ \bibnamefont {Clark}},
  \ and\ \bibinfo {author} {\bibfnamefont {D.}~\bibnamefont {Jaksch}},\ }\href
  {\doibase 10.1088/1742-5468/2013/07/p07007} {\bibfield  {journal} {\bibinfo
  {journal} {Journal of Statistical Mechanics: Theory and Experiment}\ }\textbf
  {\bibinfo {volume} {2013}},\ \bibinfo {pages} {P07007} (\bibinfo {year}
  {2013}{\natexlab{b}})}\BibitemShut {NoStop}%
\bibitem [{\citenamefont {Mendoza-Arenas}\ \emph
  {et~al.}(2013{\natexlab{c}})\citenamefont {Mendoza-Arenas}, \citenamefont
  {Al-Assam}, \citenamefont {Clark},\ and\ \citenamefont
  {Jaksch}}]{Mendoza-Arenas2013}%
  \BibitemOpen
  \bibfield  {author} {\bibinfo {author} {\bibfnamefont {J.~J.}\ \bibnamefont
  {Mendoza-Arenas}}, \bibinfo {author} {\bibfnamefont {S.}~\bibnamefont
  {Al-Assam}}, \bibinfo {author} {\bibfnamefont {S.~R.}\ \bibnamefont {Clark}},
  \ and\ \bibinfo {author} {\bibfnamefont {D.}~\bibnamefont {Jaksch}},\ }\href
  {\doibase 10.1088/1742-5468/2013/07/P07007} {\bibfield  {journal} {\bibinfo
  {journal} {Journal of Statistical Mechanics: Theory and Experiment}\ }\textbf
  {\bibinfo {volume} {2013}},\ \bibinfo {pages} {P07007} (\bibinfo {year}
  {2013}{\natexlab{c}})}\BibitemShut {NoStop}%
\bibitem [{\citenamefont {Syassen}\ \emph {et~al.}(2008)\citenamefont
  {Syassen}, \citenamefont {Bauer}, \citenamefont {Lettner}, \citenamefont
  {Volz}, \citenamefont {Dietze}, \citenamefont {Garcia-Ripoll}, \citenamefont
  {Cirac}, \citenamefont {Rempe},\ and\ \citenamefont
  {D{\"u}rr}}]{syassen2008strong}%
  \BibitemOpen
  \bibfield  {author} {\bibinfo {author} {\bibfnamefont {N.}~\bibnamefont
  {Syassen}}, \bibinfo {author} {\bibfnamefont {D.~M.}\ \bibnamefont {Bauer}},
  \bibinfo {author} {\bibfnamefont {M.}~\bibnamefont {Lettner}}, \bibinfo
  {author} {\bibfnamefont {T.}~\bibnamefont {Volz}}, \bibinfo {author}
  {\bibfnamefont {D.}~\bibnamefont {Dietze}}, \bibinfo {author} {\bibfnamefont
  {J.~J.}\ \bibnamefont {Garcia-Ripoll}}, \bibinfo {author} {\bibfnamefont
  {J.~I.}\ \bibnamefont {Cirac}}, \bibinfo {author} {\bibfnamefont
  {G.}~\bibnamefont {Rempe}}, \ and\ \bibinfo {author} {\bibfnamefont
  {S.}~\bibnamefont {D{\"u}rr}},\ }\href {\doibase 10.1126/science.1155309}
  {\bibfield  {journal} {\bibinfo  {journal} {Science}\ }\textbf {\bibinfo
  {volume} {320}},\ \bibinfo {pages} {1329} (\bibinfo {year}
  {2008})}\BibitemShut {NoStop}%
\bibitem [{\citenamefont {Garc{\'\i}a-Ripoll}\ \emph
  {et~al.}(2009)\citenamefont {Garc{\'\i}a-Ripoll}, \citenamefont {D{\"u}rr},
  \citenamefont {Syassen}, \citenamefont {Bauer}, \citenamefont {Lettner},
  \citenamefont {Rempe},\ and\ \citenamefont {Cirac}}]{garcia2009dissipation}%
  \BibitemOpen
  \bibfield  {author} {\bibinfo {author} {\bibfnamefont {J.~J.}\ \bibnamefont
  {Garc{\'\i}a-Ripoll}}, \bibinfo {author} {\bibfnamefont {S.}~\bibnamefont
  {D{\"u}rr}}, \bibinfo {author} {\bibfnamefont {N.}~\bibnamefont {Syassen}},
  \bibinfo {author} {\bibfnamefont {D.~M.}\ \bibnamefont {Bauer}}, \bibinfo
  {author} {\bibfnamefont {M.}~\bibnamefont {Lettner}}, \bibinfo {author}
  {\bibfnamefont {G.}~\bibnamefont {Rempe}}, \ and\ \bibinfo {author}
  {\bibfnamefont {J.~I.}\ \bibnamefont {Cirac}},\ }\href {\doibase
  10.1088/1367-2630/11/1/013053} {\bibfield  {journal} {\bibinfo  {journal}
  {New Journal of Physics}\ }\textbf {\bibinfo {volume} {11}},\ \bibinfo
  {pages} {013053} (\bibinfo {year} {2009})}\BibitemShut {NoStop}%
\bibitem [{\citenamefont {Poletti}\ \emph {et~al.}(2012)\citenamefont
  {Poletti}, \citenamefont {Bernier}, \citenamefont {Georges},\ and\
  \citenamefont {Kollath}}]{PolettiKollath2012}%
  \BibitemOpen
  \bibfield  {author} {\bibinfo {author} {\bibfnamefont {D.}~\bibnamefont
  {Poletti}}, \bibinfo {author} {\bibfnamefont {J.-S.}\ \bibnamefont
  {Bernier}}, \bibinfo {author} {\bibfnamefont {A.}~\bibnamefont {Georges}}, \
  and\ \bibinfo {author} {\bibfnamefont {C.}~\bibnamefont {Kollath}},\ }\href
  {\doibase 10.1103/PhysRevLett.109.045302} {\bibfield  {journal} {\bibinfo
  {journal} {Phys. Rev. Lett.}\ }\textbf {\bibinfo {volume} {109}},\ \bibinfo
  {pages} {045302} (\bibinfo {year} {2012})}\BibitemShut {NoStop}%
\bibitem [{\citenamefont {Guo}\ and\ \citenamefont
  {Poletti}(2017)}]{GuoPoletti2017b}%
  \BibitemOpen
  \bibfield  {author} {\bibinfo {author} {\bibfnamefont {C.}~\bibnamefont
  {Guo}}\ and\ \bibinfo {author} {\bibfnamefont {D.}~\bibnamefont {Poletti}},\
  }\href {\doibase 10.1103/PhysRevB.96.165409} {\bibfield  {journal} {\bibinfo
  {journal} {Phys. Rev. B}\ }\textbf {\bibinfo {volume} {96}},\ \bibinfo
  {pages} {165409} (\bibinfo {year} {2017})}\BibitemShut {NoStop}%
\bibitem [{\citenamefont {S\'a}\ \emph {et~al.}(2020)\citenamefont {S\'a},
  \citenamefont {Ribeiro},\ and\ \citenamefont {Prosen}}]{SaProsen2020}%
  \BibitemOpen
  \bibfield  {author} {\bibinfo {author} {\bibfnamefont {L.}~\bibnamefont
  {S\'a}}, \bibinfo {author} {\bibfnamefont {P.}~\bibnamefont {Ribeiro}}, \
  and\ \bibinfo {author} {\bibfnamefont {T.}~\bibnamefont {Prosen}},\ }\href
  {\doibase 10.1103/PhysRevX.10.021019} {\bibfield  {journal} {\bibinfo
  {journal} {Phys. Rev. X}\ }\textbf {\bibinfo {volume} {10}},\ \bibinfo
  {pages} {021019} (\bibinfo {year} {2020})}\BibitemShut {NoStop}%
\end{thebibliography}%

\end{document}